\documentclass[sigconf,letterpaper,authorversion,nonacm]{acmart}

\usepackage{amsfonts}
\usepackage{algorithmic}
\usepackage{textcomp}
\usepackage{xcolor}
\usepackage{amsthm}
\usepackage[ruled,vlined,linesnumbered]{algorithm2e}
\usepackage{float}
\usepackage{subfig}
\usepackage{caption}
\usepackage{array}
\usepackage{booktabs}
\usepackage{multirow}

\definecolor{TBcolor}{HTML}{007500}
\definecolor{TBcolor1}{HTML}{750000}

\setcopyright{none}

\newcommand{\QfGeo}{{\it QF-Geo}} 

\newcommand{\LCon}{\ell_{con}} 
\newcommand{\LCap}{\ell_{cap}}

\usepackage{enumitem}
\usepackage{lipsum}
\setlist[itemize]{leftmargin=*}
\setlist[enumerate]{leftmargin=*}

\AtBeginDocument{%
  \providecommand\BibTeX{{%
    \normalfont B\kern-0.5em{\scshape i\kern-0.25em b}\kern-0.8em\TeX}}}

\begin{document}

\title{\QfGeo:  Capacity Aware Geographic Routing \\ using Bounded Regions of Wireless Meshes}

\author{Yung-Fu Chen}
\affiliation{%
  \institution{The Ohio State University}
  \country{USA}
}
\email{chen.6655@osu.edu}

\author{Kenneth W. Parker}
\affiliation{%
  \country{USA}
}
\email{kenneth@ParkerTong.net}

\author{Anish Arora}
\affiliation{%
  \institution{The Ohio State University}
  \country{USA}
}
\email{anish@cse.ohio-state.edu}

\begin{abstract}


Routing in wireless meshes must detour around holes.  Extant routing protocols often underperform in minimally connected networks where holes are larger and more frequent.  Minimal density networks are common in practice due to deployment cost constraints, mobility dynamics, and/or adversarial jamming.  Protocols that use global search to determine optimal paths incur search overhead that limits scaling.  Conversely, protocols that use local search tend to find approximately optimal paths at higher densities due to the existence of geometrically direct routes but underperform as the connectivity lowers and regional (or global) information is required to address holes.  Designing a routing protocol to achieve high throughput-latency performance across a range of network densities, mobility, and interference dynamics remains challenging.

This paper shows that, in a probabilistic setting, bounded exploration can be leveraged to mitigate this challenge.  We show, first, that the length of shortest paths in networks with uniform random node distribution can, with high probability (whp), be bounded.  Thus, whp a shortest path may be found by limiting exploration to an elliptic region whose size is a function of the network density and the Euclidean distance between the two endpoints. Second, we propose a geographic routing protocol that achieves high reliability and throughput-latency performance by forwarding packets within an ellipse whose size is bounded in a similar manner as well as by an estimate of the available capacity.  Our protocol, \QfGeo, selects forwarding relays within the elliptic region, prioritizing those with sufficient capacity to avoid bottlenecks. Our simulation results show that \QfGeo\ achieves high goodput efficiency and reliability in both static and mobile networks across both low and high densities, at large scales, with a wide range of concurrent flows, and in the presence of adversarial jamming.

\end{abstract}




\settopmatter{printfolios=true}
\maketitle

\section{Introduction}

The adoption of multi-hop wireless mesh networks in access, backhaul, vehicular, swarm, or ad hoc contexts remained modest over the last couple of decades despite substantial R\&D. However, recent trends are contributing to growth in the adoption of these networks.  Among these trends is the growth in the number of edge nodes utilizing short-range communications---with high speed (i.e., mmWave or Sidelink) 5G communications \cite{niu2015survey, terragraph, molina2017lte} or low speed (i.e., LoRa or NB-IoT) communications in resource-limited and infrastructure-limited settings \cite{sinha2017survey}.  The increased availability of SDR (Software-defined Radio) (e.g., 5G-New Radio \cite{5GNR}) is contributing to this trend by providing greater interchangeability.  Likewise, flexibility in the higher levels of the networking stacks (using, e.g., OpenFlow \cite{mckeown2008openflow}) is contributing to this trend.  Finally, most commercial networks, both realized and proposed, are full of fiber nearer the core of the network, but this architecture is not easily mimicked in military applications, resulting in an acute need for mesh technologies \cite{tsm, halford2010barrage}.

These trends motivate a reconsideration of gaps that remain in the performance of multi-hop wireless mesh systems.  We will focus on the gaps in goodput efficiency, reliability, latency, and tolerance to external as well as internal interference.  In this paper, we address these gaps in the context of routing.  We are particularly interested in the quick establishment of high goodput efficiency routes, even as the networks scale in size, in the presence of limited network density and/or network dynamics---whether due to node mobility, traffic load that is variable, or interference that is benign or adversarial.  To that end, the proposed solution needs to generalize across a wide range of network configurations, in contrast to many extant routing protocols whose performance is tuned to specific configurations, e.g., high density or low interference levels.

\subsection{Challenges}

We propose that one of the primary unresolved routing challenges for wireless meshes is the difference in handling the complexity of sparse network connectivity.  This is not intended to include the issues where some of the nodes are not connected to the bulk of the network.  In ad hoc networks, there is a non-zero probability that a few nodes will be  disconnected at any density, but as the density increases beyond a critical point, the degree of disconnection falls off exponentially.  In mobile networks, such loss of connection can be short-lived and be addressed with a modicum of delay tolerance.

Define the network density, $\rho$, as the expected number of nodes within a single hop range of an arbitrary location.  Then for networks of non-trivial size, there is a critical density, $\rho_0$, such that as the density falls below $\rho_0$, the network rapidly becomes mostly disconnected, and as the density rises above this threshold, islands of local connectivity rapidly coalesce to include almost all of the nodes.%
\footnote{
    A theoretical derivation of $\rho_0$ is not known.  At one time, we was thought that $\rho_0 = \sqrt{2}$.  Numerical simulations have shown that $\rho_0$ is slightly larger than $\sqrt{2}$, but have so far failed to disprove a new conjecture $\rho_0 = \sqrt[3]{3}$.  In the remainder of the document, when the density is $\sqrt{2}$, this is intended to represent the approximate critical density.}
The transition is sharper for larger networks but never a step function (i.e., never a true phase transition).

\begin{figure}
    \includegraphics[width=0.22\textwidth]{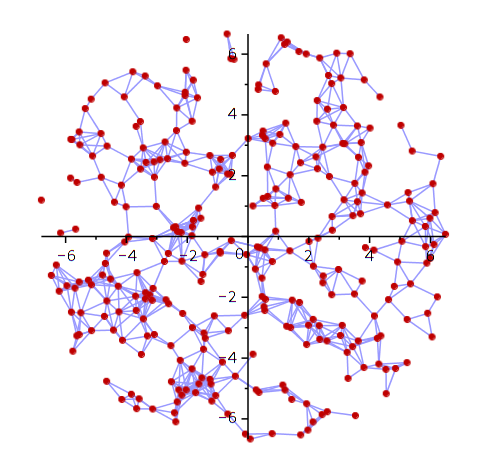}
    \hfil
    \includegraphics[width=0.22\textwidth]{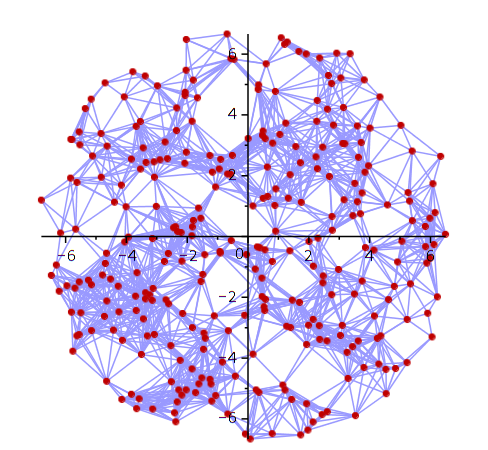}
    \caption{
        Foam-like networks vs fog-like networks.  Each figure has the same~300 randomly placed nodes.  On the left the transmission radius ($R$) is equal to 1 yielding $\rho = 2$, while on the right $R = 2$ yielding $\rho = 8$.}
    \label{fig:foamVsFog}
\end{figure}

When the network density is only slightly above $\rho_0$, there is a high probability of connectivity between any pairs of nodes, but the topology of the network is {\em qualitatively} dissimilar from  networks of high densities (e.g., $\rho\! =\! 5$).  This is illustrated by Fig.~\ref{fig:foamVsFog}.  Low-density networks have a predominance of ``holes'' where flow across the region requires substantial detours.  At higher densities, these holes largely disappear.  To make this distinction evocative, we will refer to minimally connected networks as {\em foam-like} (i.e., $\sim \!\! \sqrt{2} \leq \rho < \ \sim \!\! 2$) and highly connected networks as {\em fog-like} (i.e., $\sim\! 4 \leq \rho$).

This qualitative difference between foam-like and fog-like networks affects many existing protocols.  For example, geographic routing for point-to-point flows performs efficiently in fog-like networks where optimal routes tend to be geometrically direct; decisions based only on local information are sufficient and require minimal overhead.  But they tend to perform poorly in foam-like networks, where they backtrack upon encountering holes in the network.  The limited use of regional or global information can make a few backtrack events very inefficient, significantly distorting the average.  Conversely, many proactive and reactive routing protocols are prone to performing a global search, even when it is not necessary due to high network connectivity.  Decisions based on global search (or search over large regions) limit the network scaling, because of high control overhead and/or long latency.

Another primary challenge is related to the stability of paths, which is decreased by node mobility, changes in external interference (including adversarial jamming), and even changing traffic patterns.  Because these problems are exacerbated by increasing network scale, and because the speed at which the network can repair itself (or refine routing paths) slows with increasing scale, at some scale, these problems lead to network failure \cite{kulathumani2016repair}.  Moreover, routing in some applications has to account for the network transitioning from foam-like to fog-like, or vice versa.

\subsection{Approach}

Foam-like networks maintain connectivity, despite frequent and larger holes, through indirect routes.  Let us define {\em path stretch} as the ratio of the length of the shortest path to the Euclidean distance between the endpoints.  Our experimental results demonstrate that path stretch can be over 10x for short paths at minimal network densities and that path stretch decreases for longer paths and at higher network densities.  Quantifying this trend allows a probabilistic bound on the path length.  For example, at a density of 2, there is a 99\% chance that endpoints that are 5~transmission units apart have a path stretch of less than 3.1, implying that 99\% of the time, the length of a shortest path between the end points (at this network density) will be less than 15.5 transmission radius.



Let $\mathbb{P}$ denote the probability of finding a shortest path in a defined region. By using regression over many simulations of uniform random networks and random pairs of endpoints, a function was derived for the smallest ellipse that contains a shortest path between a given pair of endpoints with probability $\mathbb{P}$. This function depends both on the distance between the endpoints and the density of the network.  Only ellipses with vertices at the source and the destination are considered.  For notational convenience, let $\LCon$ be the ratio of the length of the major axes of this ellipse to the distance between the endpoints.  The basic approach to routing in a network, given a source and a destination, is to search for a shortest path within the elliptic region determined by the function.

This approach to routing is refined to accommodate concurrent flows and external interference in the network, as follows:
%
%
%
\vspace*{-1mm}
\begin{enumerate}
    \item Estimate an effective density $\rho'$ ($\leq \rho$) of the network, which is the density resulting from only using the capacity not already allocated to other flows. Accordingly, determine an ellipse factor, $\LCap$ ($\geq \LCon$), for a larger ellipse that whp contains a shortest path in the presence of internal and external interference.
    
    
    \item Search for a shortest path that avoids exceeding the capacity of relay nodes where possible.
    
    
    \item Track the residual capacity at each node for use in the process.
\end{enumerate}

\vspace*{-2mm}
\subsection{Contributions of the Paper}

This paper shows the feasibility of a probabilistic framework that limits exploration while yielding a sufficiently high probability of routing success when paths exist. This is in contrast to routing based on worst-case analysis, which leads to an overly pessimistic framing of the solution space and incurs higher overhead and lower stability.  

Secondly, an elliptic search region is computed to serve as a slightly conservative but reasonably tight estimate of the region for exploration. The result can be directly applied without modification to many networks that only approximate the uniform constant density assumption. For classes of networks that  depart radically from these assumptions, it is our position that the deployment distribution information can and should be incorporated into the routing system, in an analogous manner.

Thirdly, a design of geographic routing protocol, \QfGeo\ (for QuickFire-Geo), is presented.  \QfGeo\ is based on a depth-first search in the elliptic region.  \QfGeo\ offers notably improved performance in foam-like networks, where ``optimistic'' routing approaches almost always underperform, often dramatically.  \QfGeo\ also performs well in fog-like networks where ``conservative'' algorithms, especially proactive worst case algorithms, often incur overhead that is rarely needed.  That is \QfGeo\ yields nearly optimal goodput efficiency across a wide range of densities.  Almost all existing algorithms make implicit assumptions about network topology that favor either fog-like or foam-like networks and ensure dramatic loss of efficiency when these assumptions do not hold.


This combination of approaches leads to a useful middle ground between aggressively proactive and highly reactive approaches.  If changes to the network, its environment, or the traffic are infrequent a proactive algorithm that approximates maintenance of all relevant network information may be acceptable.  But in more dynamic scenarios and at larger scales it becomes impossible to maintain so much information.  In contrast an extremely reactive approach makes finding a route on demand quite expensive, which also limits scaling.  We postulate that a location service is approximately the minimal proactive infrastructure that doesn't result in acute on-demand route discovery problems.  However, having found a route with minimal proactive infrastructure \QfGeo\ proactively maintains it for the duration of the flow.

%
Our simulation results for \QfGeo\ show:
\vspace*{-1mm}
\begin{itemize}
    \item \QfGeo\ achieves high goodput efficiency and has a high likelihood of finding reliable paths, notably in diverse network configurations as well as in the presence of changes in network density, node mobility, external interference, and jamming. 
    
    \item \QfGeo\ achieves its improved performance due to limiting the search region and its introduction of memory of the routing path. 

    \item Proper selection of the ellipse factor improves the goodput efficiency by reducing the latency while not harming the reliability.   
    
    \item The memory of the last forwarder in \QfGeo\ allows ongoing flows to further reduce their packet traversal time and optimize the packet reception reliability, especially when nodes are mobile.  Moreover, memory improves the goodput efficiency in scenarios of low-density networks with and without mobility as well as in high-density networks with mobility.
\end{itemize}

Our simulations are comparative with Geographic Routing and with Maximum Capacity Routing 
\cite{alwan2014performance}. Extensive simulations comparing with other state-of-the-practice protocols such as Optimized Link State Routing (OLSR) \cite{clausen2003optimized} and Ad hoc On Demand Distance Vector (AODV) \cite{perkins2003ad} show even greater improvements in \QfGeo; these are, however, omitted from this paper for lack of space. 

\subsection{Organization of this Paper}

In Section 2, we discuss \QfGeo\ in the context of related state-of-the-art routing solutions in wireless mesh networks. In Section 3, we outline our numerical analysis of the path stretch and the existence of tight bounds on the minimum search region for the shortest path. In Section 4, we present the \QfGeo\ routing protocol in terms of a detailed explanation of its distributed approach for the bounded search of route forwarding. We describe our evaluation and simulation results, 
and make concluding remarks in Section 6. 

\section{Related Work}



Proactive algorithms tend to require global information \cite{boukerche2011routing}, while reactive algorithms tend to invoke global search \cite{alotaibi2012survey}.  The extent to which non-local information is used or non-local activity is required is a primary limitation on scaling.  As an aside, scaling is only challenging in the presence of changes in the network topology, the traffic, or the channel.  As networks scale up, proactive algorithms tend to incur high control overhead, and reactive algorithms tend to have high setup latency.  

Back pressure style algorithms, \cite{moeller2010routing, ji2012delay}, can asymptotically approach the network's achievable capacity at the expense of substantial packet-level delays.  However, trading substantial increases in latency for often modest  increases in aggregate throughput is rarely desired by applications.  And even if this trade-off is acceptable, back pressure algorithms tend to respond slowly to changes in the network topology (unless those changes can be problematically characterized in advance).  As a result, back pressure style algorithms are rarely deployed.


Geographic routing protocols leverage geometric location information to forward data with low overhead \cite{cadger2012survey}. As a simple policy option, greedy forwarding chooses the relay geographically closest to the destination. Yet, it may become trapped on the edge of a hole without any neighbor that is closer to the destination.  The face routing algorithms \cite{ko1999geocasting, karp2000gpsr, kuhn2008algorithmic} guarantee packet delivery by constructing a planar subset of the network and then walking along the faces of this planar graph that intersect with the line between the source and destination.  Creating and maintaining a planar subset of the network is a proactive infrastructure that impacts the network capacity as the network scales.  When the holes are frequent and large, the best path is likely to involve long detours, but the cost of discovering such a path through face routing is high. Even more problematic, classical geographic routing rediscovers the route for every packet. Finally, geographic routing is typically not capacity aware, so its performance is impacted by its tendency to route too much traffic thru hot spots. For low-density networks, the throttling caused by hot spots can substantially limit performance across a large portion of the network when traffic approaches capacity.


Another scheme uses a proactive infrastructure to maintain awareness of all holes in the network and then uses geographic routing to approximately follow a quarter ellipse that skirts the known hole, \cite{yang2021ellipse}. As with all landmark-based routing schemes, arbitrary landmarks increase the packet cost, reduce aggregate capacity (by ignoring other routes), and create hot spots.  Additional problems arise when many different-sized holes appear between the source and the destination.  This use of ellipses has no connection to the use of elliptic search regions presented here.




The AFR routing protocol uses ellipses to bound a search region in a manner that can be contrasted with this paper \cite{kuhn2002asymptotically}.  It assumes that the network is a Unit Disk Graph (UDG), which is equivalent to one of the assumptions in this paper, but make no assumptions on the locations of the nodes (except that they are not arbitrarily close together, which is used to prove a lower bound on the cost of the resulting path).  This allows for highly unlikely typologies that are ``pathologically'' challenging.  They then focus on worst-case performance, which is dominated by such challenging but unlikely cases.  As a result, they prove that the worst-case complexity of finding a route is $O\left( {c_b}^2 \right)$, where $c_b$ is the cost of the best path between the source and the destination.  However, $c_b$ might be prohibitively large, e.g., $O\left( n^2 \right)$, where $n$ is the number of nodes.

In contrast, this paper assumes that the nodes are uniformly randomly distributed over the network.  This allows a much tighter probabilistic bound on $c_b$, based on the Euclidean distance between two endpoints and the network density. The resulting algorithm ensures, with arbitrary probability (i.e., 99\%), that the best path will be found at a dramatically lower cost than in AFR.  That is, the key difference is the focus on the probabilistic setting instead of the worst-case setting.

Secondary differences include:
\begin{itemize}
    \item AFR lacks foreknowledge of the appropriate ellipse size, so it iterates from smaller to larger sizes until it finds a suitably large ellipse.  In contrast, \QfGeo\ knows in advance, with useful accuracy, the appropriate size ellipse.
    
    \item AFR uses Face Routing, which requires proactive maintenance of a planar embedding of the actual network, whereas \QfGeo\ uses a depth-first search of the elliptic sufficient search region, which is more efficient than face-routing, even ignoring the issues of planar embedding.
    
    \item AFR appears to rediscover the route for every packet,  whereas \QfGeo\ discovers the route during flow setup and then remembers and updates the route during the (often short) life of the flow.
\end{itemize}

\section{Analysis of Path Stretch and Boundedness of Search Region}

\begin{figure}[!t]
\minipage{0.236\textwidth}
  \subfloat[$\rho = \sqrt{2}$]{\label{circuitousness_density_1.414}{\includegraphics[width=\linewidth]{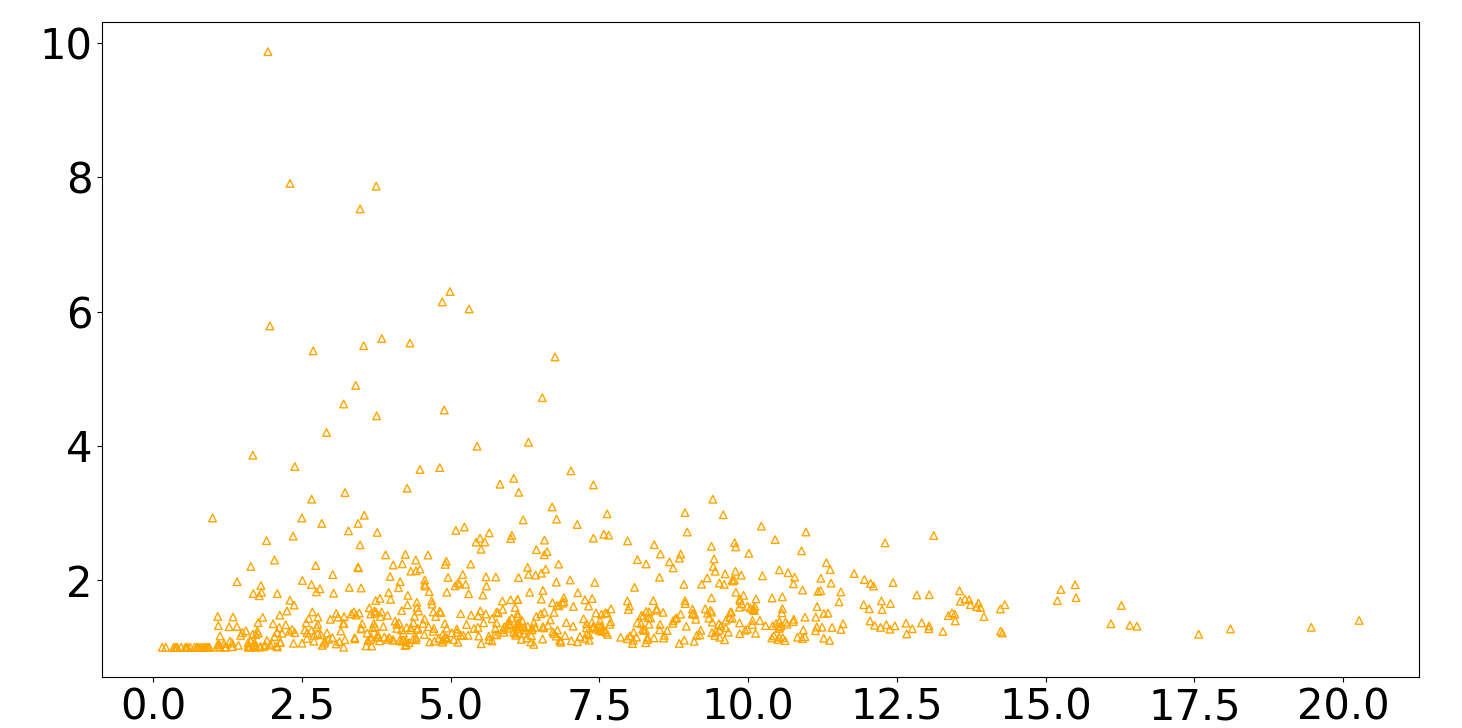}}}
\endminipage\hfill
\minipage{0.236\textwidth}
  \subfloat[$\rho = 2$]{\label{circuitousness_density_2}{\includegraphics[width=\linewidth]{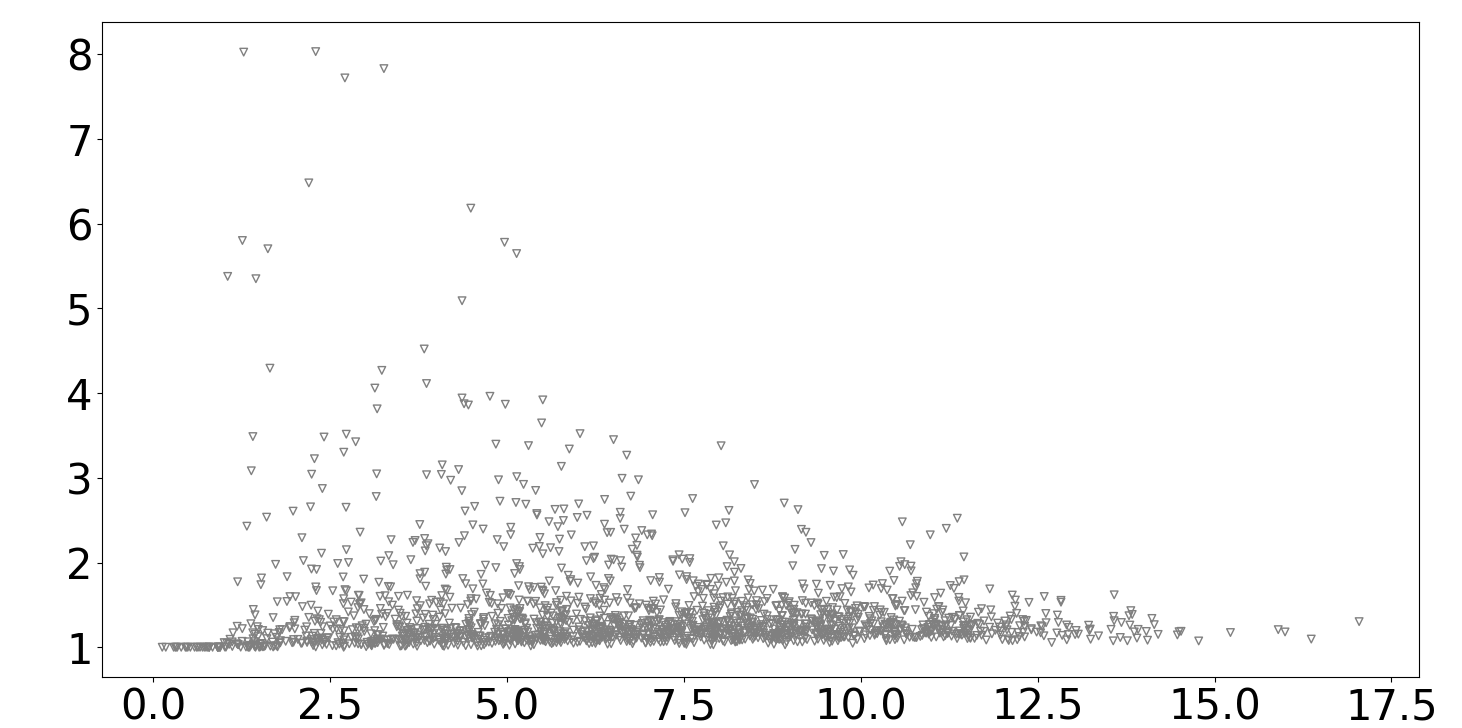}}}
\endminipage\hfill
\\
\minipage{0.236\textwidth}%
  \subfloat[$\rho = 3$]{\label{circuitousness_density_3}{\includegraphics[width=\linewidth]{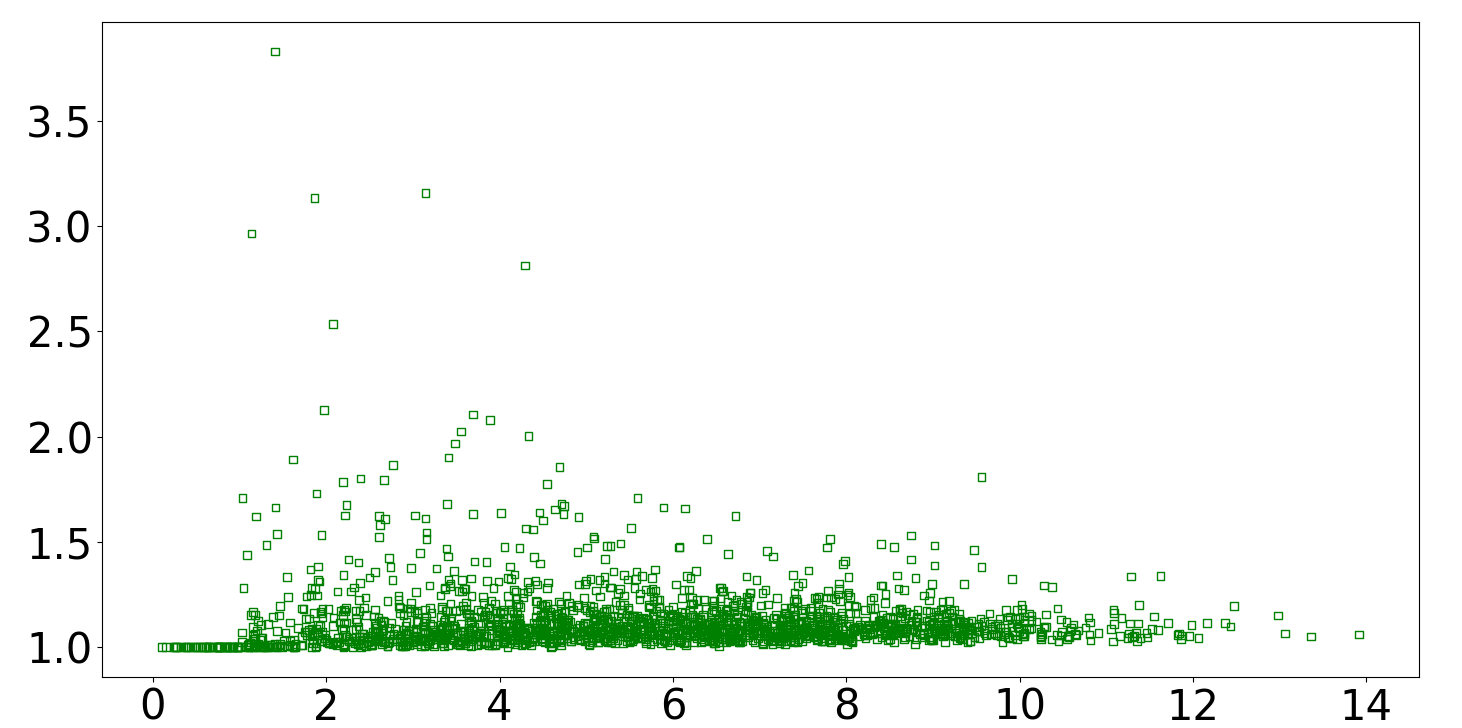}}}
\endminipage\hfill
\minipage{0.236\textwidth}%
  \subfloat[$\rho = 4$]{\label{circuitousness_density_4}{\includegraphics[width=\linewidth]{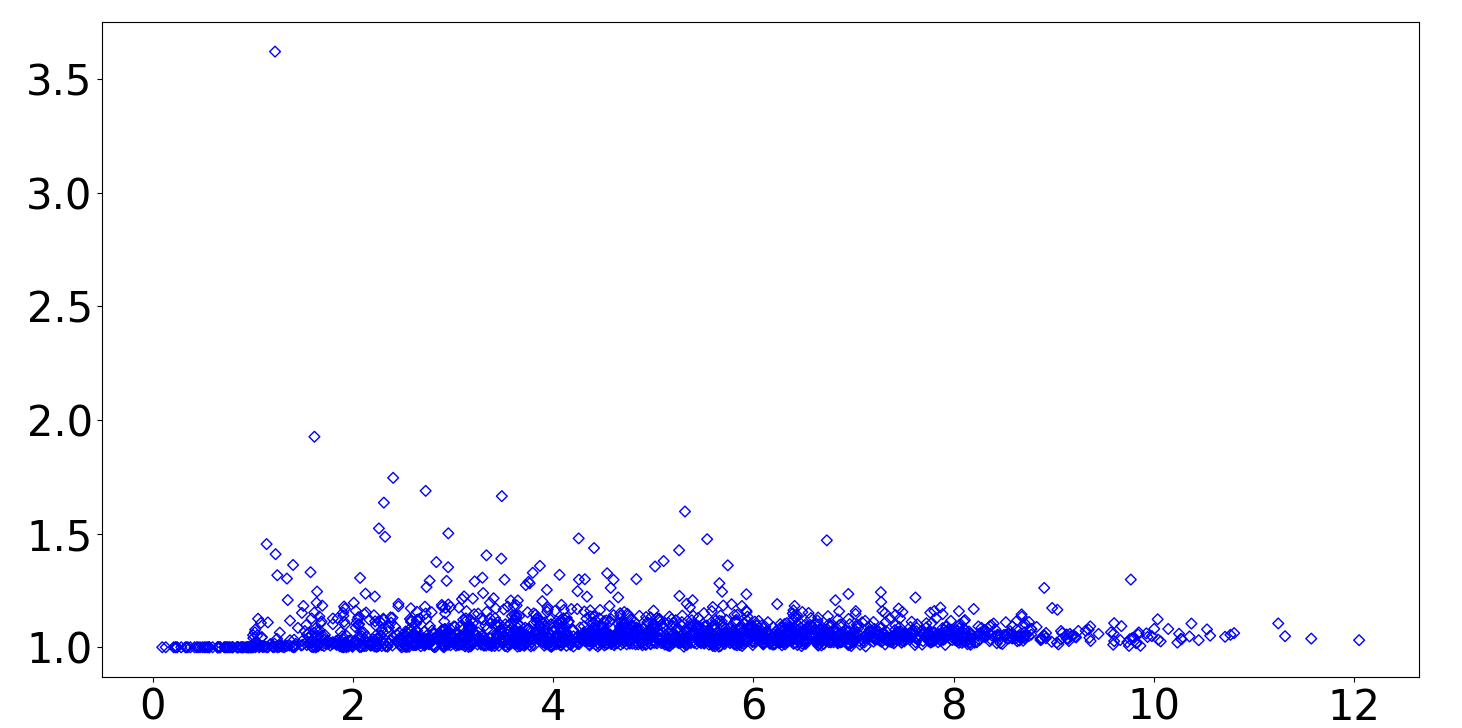}}}
\endminipage\hfill
\\
\minipage{0.236\textwidth}%
  \subfloat[$\rho = 5$]{\label{circuitousness_density_5}{\includegraphics[width=\linewidth]{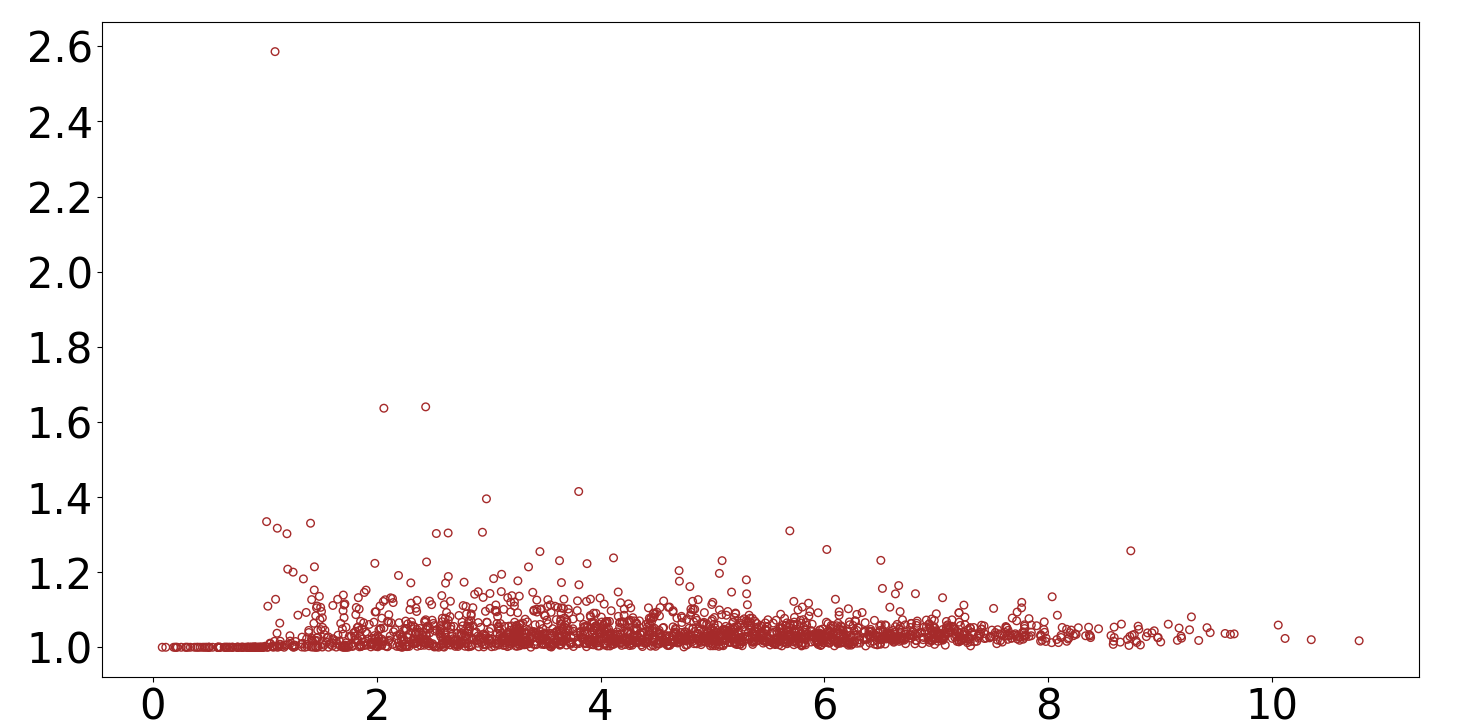}}}
\endminipage
\caption{Scatter plot of path stretch $(\delta, \zeta)$ for endpoints $(v_s, v_d)$ in 2000 random networks per configuration with $n = 343$ and density $\rho \in {\{\sqrt{2}, 2, 3, 4, 5\}}$.}
\label{circuitousness}
\end{figure}

This section describes Monte Carlo simulations that quantify the path stretch. The objective of the simulation is to demonstrate the correlation between path stretch and the network parameters in uniform random networks. Furthermore, by analyzing the simulation results, we determine a probability bound ($\mathbb{P} = 99\%$) on path stretch and thus justify exploring routing paths within an ellipse predicted by a model with only a few inputs of network parameters.

\begin{figure}[!t]
\minipage{0.236\textwidth}
  \subfloat[$\rho = \sqrt{2}$]{\label{EF_density_1.414}{\includegraphics[width=\linewidth]{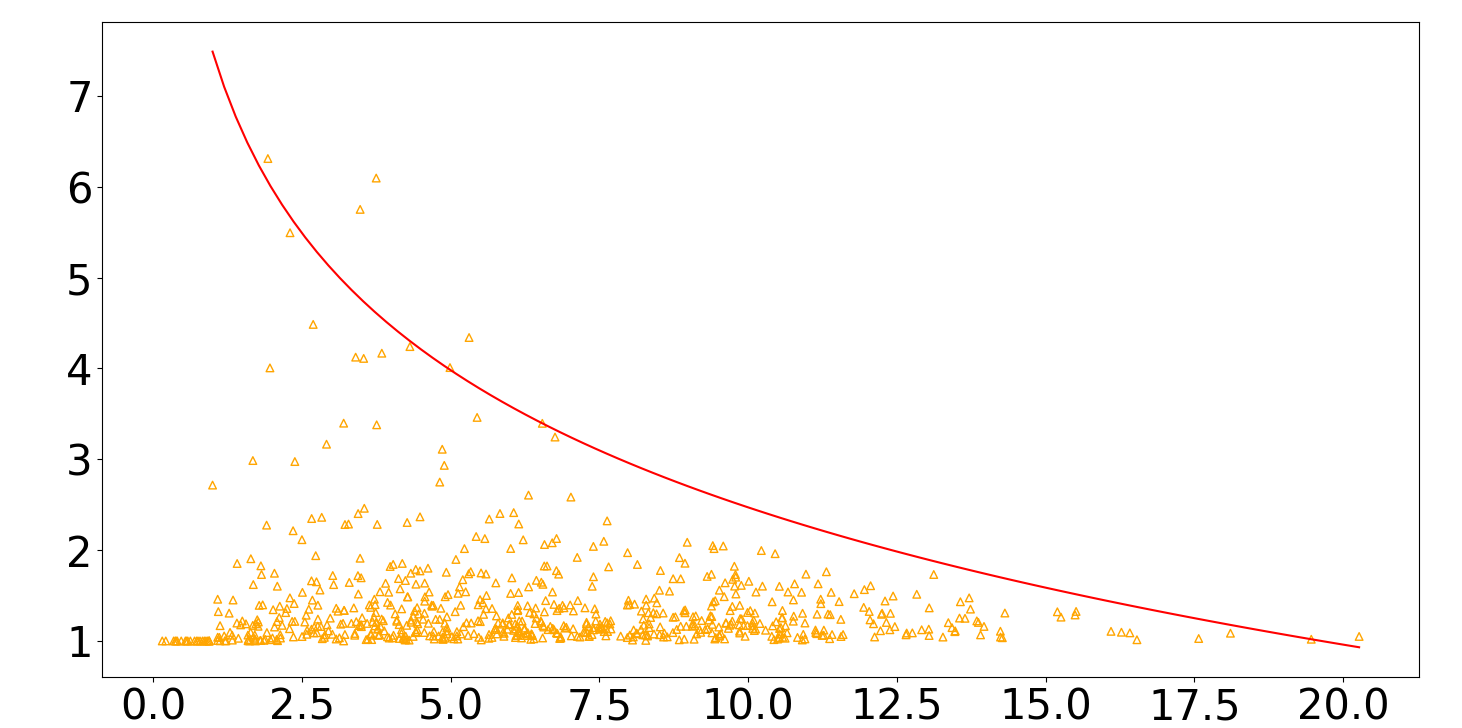}}}
\endminipage\hfill
\minipage{0.236\textwidth}
  \subfloat[$\rho = 2$]{\label{EF_density_2}{\includegraphics[width=\linewidth]{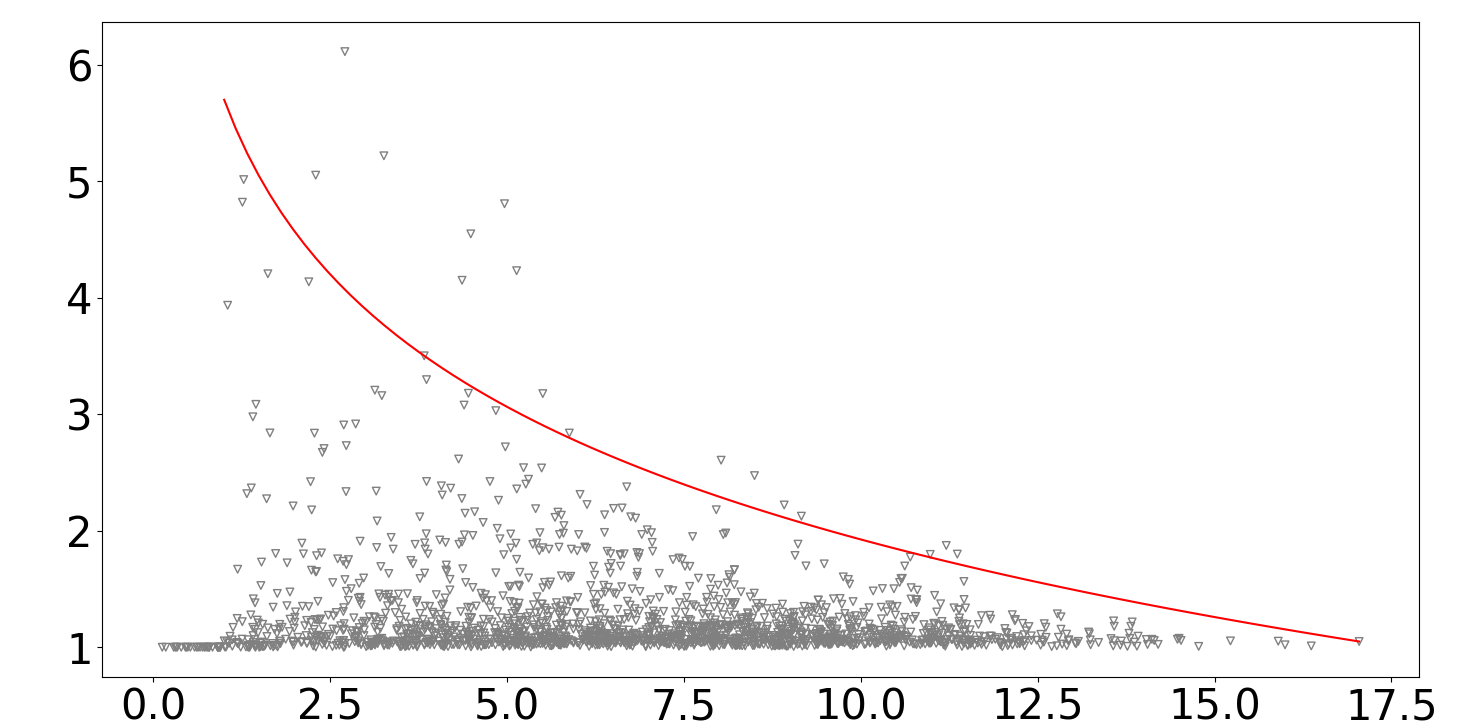}}}
\endminipage\hfill
\\
\minipage{0.236\textwidth}%
  \subfloat[$\rho = 3$]{\label{EF_density_3}{\includegraphics[width=\linewidth]{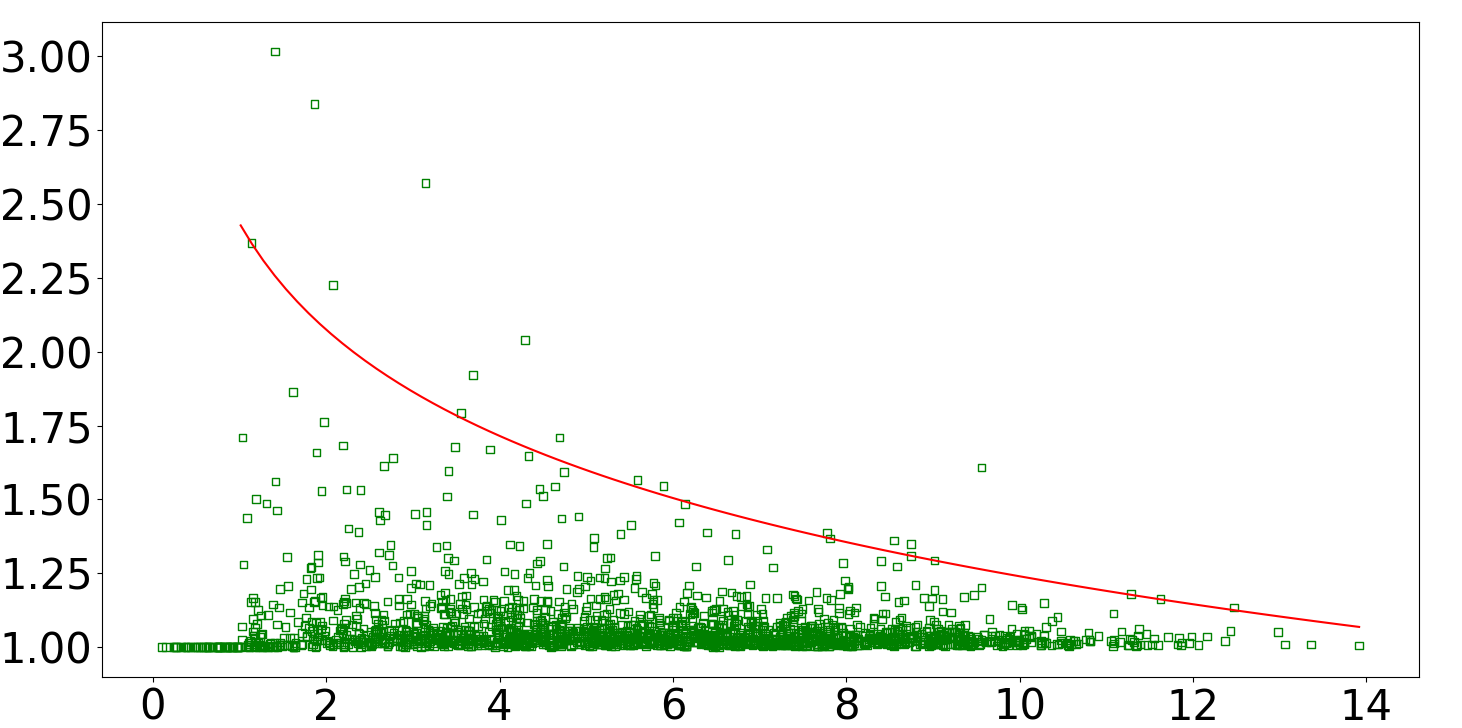}}}
\endminipage\hfill
\minipage{0.236\textwidth}%
  \subfloat[$\rho = 4$]{\label{EF_density_4}{\includegraphics[width=\linewidth]{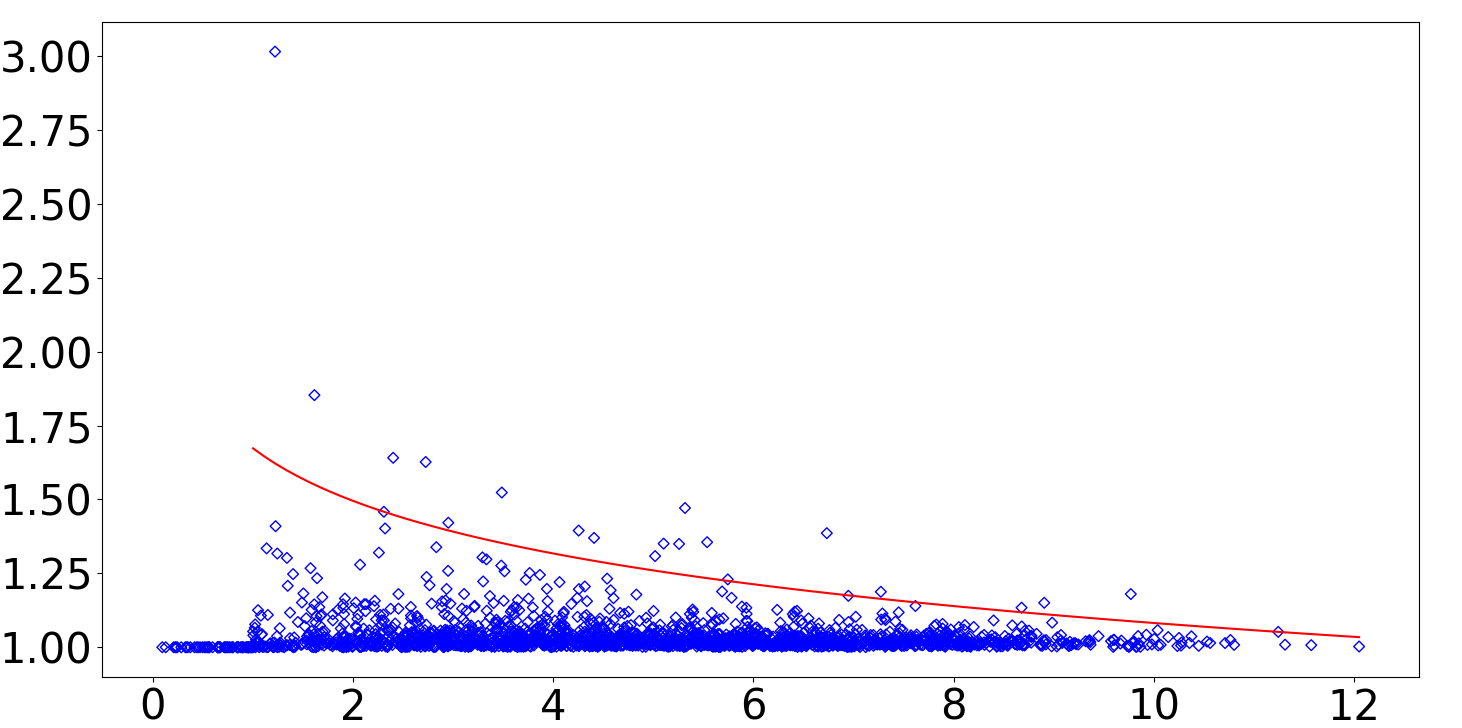}}}
\endminipage\hfill
\\
\minipage{0.236\textwidth}%
  \subfloat[$\rho = 5$]{\label{EF_density_5}{\includegraphics[width=\linewidth]{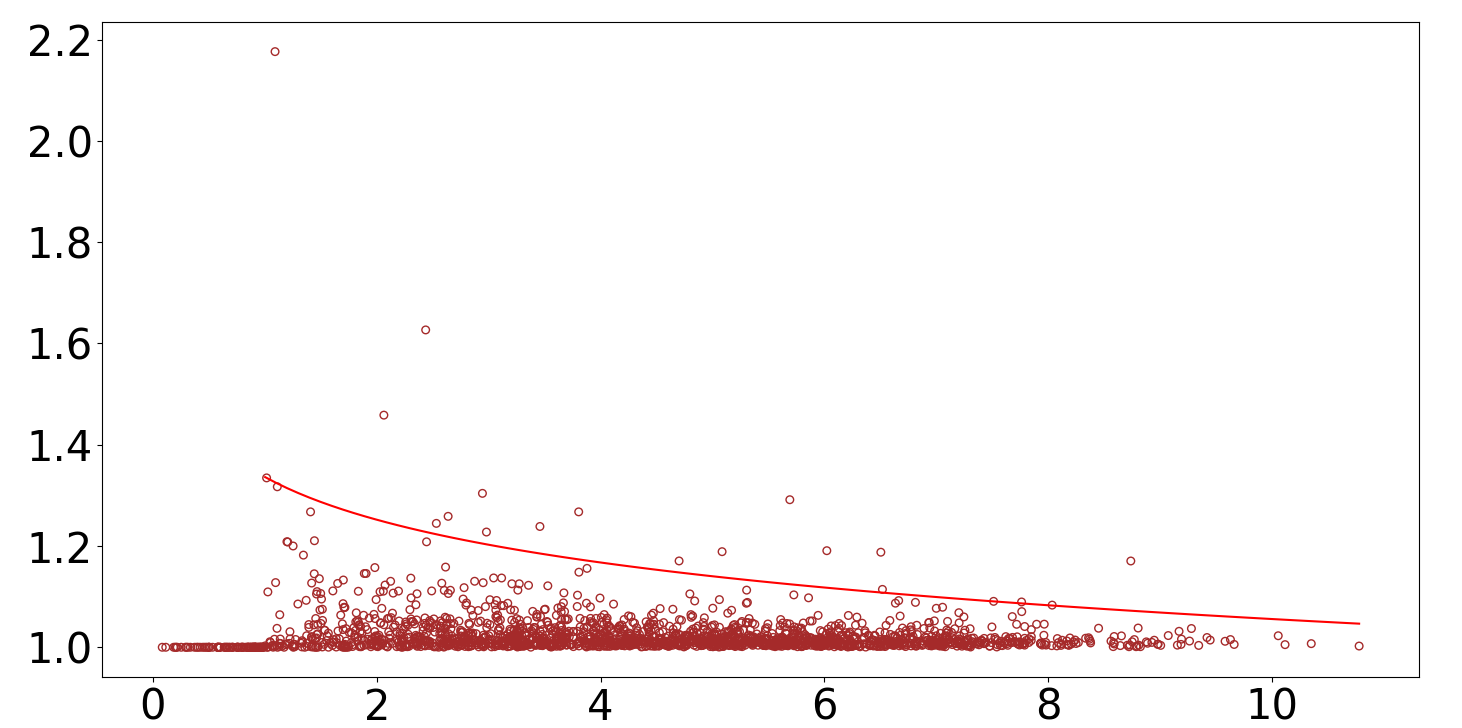}}}
\endminipage
\caption{Scatter plot of ellipse factor $(\delta, \LCon)$ for endpoints $(v_s, v_d)$ of the same network graphs as in Fig.~\ref{circuitousness}. The red curves denote the 99th percentile bound of $\LCon$ derived from quantile regression.}
\label{ellipse_factor}
\end{figure}

\subsection{Simulating the Path Stretch} \label{simulation_circuitousness}

Let $d_e$ denote the Euclidean distance between two endpoints and let $d_{sp}$ denote the length of the shortest path between the endpoints. Define unitless versions of these distances as $\delta := d_{e}⁄R$ and $s := d_{sp}⁄R$, where $R$ is the transmission radius. Further, define the path stretch as the ratio $s/\delta$ and denote it by $\zeta$. Clearly, $1 \leq \zeta$ because the path length can never be less than the Euclidean distance. And if $\delta \leq R$, then $\zeta=1$ because there is a direct link (i.e., a one-hop path) between the endpoints. Note that if two nodes are not connected, then $s \! = \! \infty$ and $\zeta \! = \! \infty$. Since the probability of at least some network nodes being disconnected is non-zero, in the statistics, we only count those networks for which there exists at least one route between the given two endpoints.

To generate samples from a 2D distribution of $(\delta, \zeta)$, we construct a network with node locations of uniform random distribution and links determined by geometric proximity. Let $n$ denote the number of nodes, and $\rho$ denote the network density with the definition of the average number of nodes per $R^2$ area. Then all nodes are distributed in a square with the side $w = \sqrt{\frac{n \times R^2}{\rho}}$. Our analysis is based on 2000 networks per configuration with $n \! = \! 343$ and $\rho \! \in \! {\{\sqrt{2}, 2, 3, 4, 5\}}$. We select a fixed pair of endpoints $(v_s, v_d)$ in each example and compute the shortest paths between these two endpoints accordingly. Since the examples without connectivity between the selected endpoints are filtered out, the number of networks per configuration we count in the analysis may vary and be less than 2000.

\subsection{Distribution of Path Stretch}
Fig.~\ref{circuitousness} plots the path stretch across different network densities. The subfigures show that the variance becomes more significant as $\delta$ decreases. Also, given a value of $\delta$, the corresponding average $\zeta$ reduces as the network density decreases. Notice for short connections that, while there is a substantial chance of less than a 1\% net path stretch, there is also a substantial chance of path stretch that ranges from up to 10x to up to 2.6x as the network density ranges over ${\{\sqrt{2}, 2, 3, 4, 5\}}$.  These observations imply that, with high probability, there is a tight bound on path stretch $\zeta$, and thereby a shortest path exists in a limited network region that is determined by $\rho$ and $\delta$. We are therefore motivated to validate if bounded search regions can be associated that whp contain a shortest path and for which it is feasible to predict the size of the regions in terms of the $\rho$ and $\delta$ parameters.

\subsection{Simulating the Bounded Search Region}

Given a graph and the two endpoints $(v_s, v_d)$ in it as foci, the ellipse factor $\LCon$ can be defined to shape the ellipse whose major axis is $\LCon \times \overline{v_s v_d}$. For a path $pt$, we define $V_{pt}$ as the set of nodes on path $pt$. Then the ellipse factor in determining the minimum-sized ellipse to include $pt$ can be obtained as follows:
\begin{equation}
\begin{array}{l}
\LCon = \frac{max(\overline{v_s v_r}+\overline{v_r v_d})}{\overline{v_s v_d}}, v_r \in V_{pt}
\label{eq_ellipse_facotr_connectivity}
\end{array}
\end{equation} 

To simulate the distribution of the bounded search region across different $\rho$ and $\delta$, we calculate $\LCon$ from the shortest path dataset in Fig.~\ref{circuitousness}.

\subsection{The Ellipse Factor for Achieving Connectivity}

Fig.~\ref{ellipse_factor} shows the distribution of $\LCon$ respectively for densities ranging over $\rho \in {\{ \sqrt{2}, 2, 3, 4, 5\} }$. The red curve in each subfigure represents the 99th percentile bound of the ellipse factor; the bound is computed using quantile regression \cite{koenker2001quantile} in a logarithmic scale with a model of $w_1 \ln{\delta}+w_0$ over the weights $w_1$ and $w_0$. The curves show bounds on the $\LCon$ such that the shortest path can whp be explored in a limited region determined by $\rho$ and $\delta$. Note that the ellipse with the factor of $\LCon$ may include paths $pt$ longer than the shortest path $sp$, whose corresponding ellipse factor $\ell_p$ is less than $\LCon$, but searching in those smaller ellipses would not be desirable as it would likely yield a suboptimal route. The search region should be the minimum subgraph that includes the shortest path.

\subsection{The Model for Prediction of $\LCon$}
\label{prediction_l_connectivity}

Our objective is to predict the minimum search region of the shortest path to route packets according to the network density and source-destination distance. To that end, we propose the following model to calculate $\LCon$ given $\rho$ and $\delta$:
\begin{equation}
\begin{array}{l}
\LCon(\rho, \delta) = \begin{cases}
1,\ i\!f \ \delta \leq 1 \\
max(1+\frac{\alpha \ln{\delta}+\beta}{\rho^\gamma}, \ell_{min}),\ otherwise ,
\end{cases}
\label{eq_prediction}
\end{array}
\end{equation} 
where $\alpha$, $\beta$, and $\gamma$ are coefficients that determine the size of an ellipse in accordance with the network parameters (i.e., $\rho$ and $\delta$). The lower bounds of $\LCon$, 1 and $\ell_{min}$, are respectively provided for $\delta \leq 1$ and $\delta > 1$.

\begin{figure}[hbt!]
  \centerline{\includegraphics[width=0.48\textwidth]{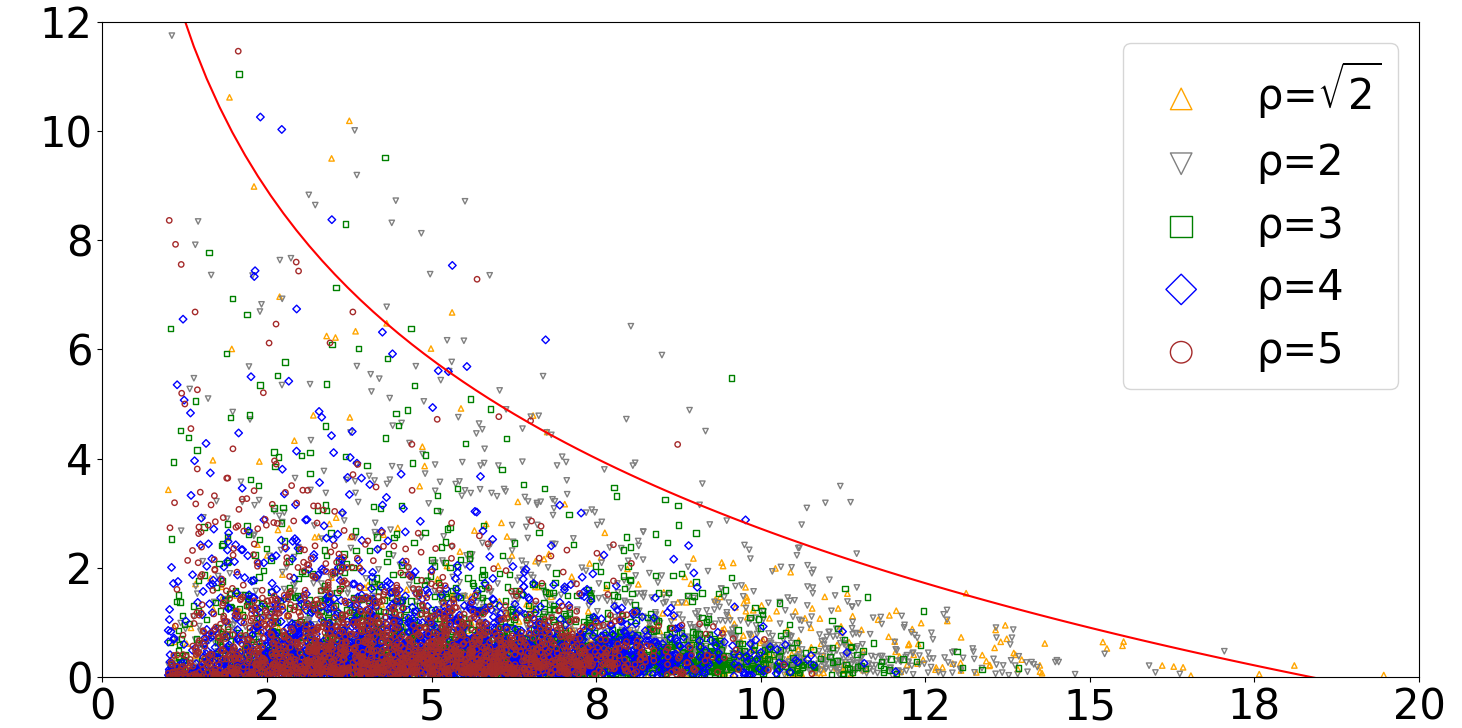}}
  \caption{Scatter plot of normalized ellipse factor $(\delta, \hat{\ell}_{con})$. The red curve denotes the 99th percentile bound of $\hat{\ell}_{con}$ derived from quantile regression.}
  \label{circuitousness_normalized}
\end{figure}

The parameters of the model are calculated as follows. First, we determine the value of $\gamma$ such that normalized values of $\LCon$ ($\hat{\ell}_{con} = (\LCon - 1)* \rho^\gamma$) in the training dataset have a similar distribution over the y-axis across different densities. Thus, we choose $\gamma = 2$; the corresponding distribution of $(\delta, \hat{\ell}_{con})$ is plotted in Fig.~\ref{circuitousness_normalized}. Next, we compute $\alpha$ and $\beta$ using quantile regression at the 99th percentile using a standard logarithmic model, $\hat{\ell}_{con} = \alpha \ln{\delta}+\beta$. The red curve in Fig.~\ref{circuitousness_normalized} represents the 99th percentile bound of normalized ellipse factor $\hat{\ell}_{con}$. Table~\ref{notationQuickFire} shows for each $\rho \! \in \! {\{ \sqrt{2}, 2, 3, 4, 5\} }$ the percentage of the ellipse factors $\LCon$, which are predicted by Eq.~\ref{eq_prediction}, that are below the red curves in Fig.~\ref{ellipse_factor}; in each case, the percentage is close to the targeted 99\%.



\begin{table}[htb!]
\caption{Percentage of values of $\LCon$, predicted by Eq.~\ref{eq_prediction} with $\alpha=-4.4732$, $\beta=13.0715$, $\gamma=2$, $\ell_{min}=1.05$, that are below the 99th percentile bound for $\rho \! \in \! {\{ \sqrt{2}, 2, 3, 4, 5\} }$ in Fig.~\ref{ellipse_factor}.}
\begin{center}
\begin{tabular}{|c|c|c|c|c|}
\hline
\textbf{$\rho = \sqrt{2}$} & \textbf{$\rho = 2$} & \textbf{$\rho = 3$} & \textbf{$\rho = 4$} & \textbf{$\rho = 5$}\\
\hline
98.53\% & 97.27\% & 99.59\% & 99.48\% & 98.69\%\\
\hline
\end{tabular}
\label{notationQuickFire}
\end{center}
\end{table}

\section{The \QfGeo\ Protocol}

\subsection{Overview of \QfGeo}

\QfGeo\ performs per-packet routing between two endpoints $(v_s, v_d)$ within an ellipse computed by a function of network density, source-destination distance, and current traffic. To predict the minimum size of the ellipse, which offers requirements for connectivity (i.e., a subgraph including the shortest path) and capacity (i.e., a subgraph including sufficient residual capacity), \QfGeo\ respectively calculates $\LCon$ and $\LCap$. The $\LCon$, predicted by the model in Eq.~\ref{eq_prediction}, is the ellipse factor that determines a search region in which nodes $v_r$ included in a shortest path, $\overline{v_s v_r}+\overline{v_r v_d} \leq \LCon\overline{v_s v_d}$, can be found with probability $\mathbb{P}$. In the presence of concurrent communication and external interference, we calculate another ellipse factor $\LCap$, $\LCap \geq \LCon$, to heuristically approximate a minimum area that provides for sufficiently many forwarders whose residual capacity $\Theta$ satisfies the capacity requirement $C$.


\QfGeo\ builds on the inherent local search in geographic routing. Recall that in geographic routing, a suboptimal path is found by greedily forwarding packets to the neighbor closest to the destination in a high-density network when geometrically direct routes exist. However, in a foam-like network, backtracking may be required when encountering holes without prior knowledge. Thus, in order to ensure goodput efficiency and reliable packet delivery, \QfGeo\ applies a depth-first search scheme to traverse the subgraph within a designated elliptic area. Each packet keeps information on visited nodes to distinguish unvisited neighbors and maintains a stack of the explored path to determine the forwarder when backtracking. To further improve performance, \QfGeo\ prioritizes forwarding relay selection to avoid relays whose capacity is below a threshold \cite{kim2009interference}; it also maintains a per-flow memory of the last forwarder so as to eschew forwarding to nodes that will backtrack and thus save the routing search time. 
To accommodate potential network mobility, it also uses an $\epsilon$-greedy approach to leverage exploitation and exploration activities to adapt to the topology change. With a probability of $\epsilon$, \QfGeo\ explores a new route so as to adapt to network changes.

\subsection{The Ellipse Factor for Achieving Capacity}

For capacity-aware forwarding, each source node $v$ maintains an estimate of the density $\rho' (\leq \rho)$ of a subgraph of the network that includes only nodes whose residual capacity exceeds a capacity threshold $C_{min}$. 
To that end, it periodically estimates its residual capacity $\Theta(v)$ by observing the outgoing data traffic and interference traffic.\footnote{The residual capacity $\Theta(v)$ can be approximated by a local estimation of $max(1 - (D_{v}+e^{I(v)}), 0)$ \cite{li2010chameleon},  where $D_{v}$ is the probability that node $v$ sends out a packet and $I(v)$ is an interference estimator of the probability that some interferers transmit within $v$'s interference range.} $\Theta(v)$ is used not only to predict $\LCap$ but also as the metric used to prune forwarding candidates. Also, by collecting $\Theta(u)$ from node $u$ within $v$'s neighborhood, $v$ locally builds a distribution model to estimate the probability distribution of residual capacity in a region centered around $v$. Thus, given a capacity threshold $C_{min}$, $v$ is able to predict the percentage of nodes with a residual capacity greater than $C_{min}$, denoted by $\phi \in [0, 1]$.

Using the effective density $\rho'=\phi \rho$ and the source-destination distance, $\LCap$ is calculated by the model in Eq.~\ref{eq_prediction}:
\begin{equation}
\begin{array}{l}
\LCap(\rho, \delta, \phi) = \LCon(\phi \rho, \delta).
\label{eq_ellipse_facotr_capacity}
\end{array}
\end{equation} 

Note that underestimation of $\LCap$ may result in forwarding packets to bottlenecks and thus cause transmission failure. For a given packet that exceeds a retransmission limit, \QfGeo, therefore, allows any relay $v_r$ holding this packet to recalculate $\LCap$ used for routing between $v_r$ and $v_d$.

\subsection{The \QfGeo\ Algorithm}

\begin{table}[htb!]
\vspace*{.5mm}
\caption{Notations for QuickFire}
\begin{center}
\begin{tabular}{c|c}
\hline
\textbf{Symbol}&\textbf{Meaning} \\
\hline
$p$ & packet id\\
\hline
$\ell$ & ellipse factor\\
\hline
$src$ & source node id\\
\hline
$dst$ & destination node id\\
\hline
$f\!l$ & flow id\\
\hline
$\rho'$ & effective network density\\
\hline
$\Theta(v)$ & estimated residual capacity of node $v$\\
\hline
$C(f\!l)$ & capacity requirement of flow $f\!l$\\
\hline
$RETX(p, v)$ & retransmission times of packet $p$ at node $v$\\
\hline
$RETX_{max}$ & packet retransmission limit at a forwarder\\
\hline
$BL(p, v)$& set of neighbors backtracking to $v$ for packet $p$\\
\hline
$q(f\!l, v)$ & last forwarding target at $v$ in flow $f\!l$\\
\hline
$N\!E(p, v, \ell)$ & set of $v$'s neighbors \\
& within the forwarding ellipse of $p$ \\
\hline
$Dist(v, u)$ & Euclidean distance between nodes $v$ and $u$\\
\hline
$Parent(p, v)$ & parent of node $v$ in packet $p$'s traversed path\\
\hline
$V(p)$ & set of visited nodes for packet $p$\\
\hline
$rnd$ & random value $\in$ [0, 1]\\
\hline
\end{tabular}
\label{notationQFGeo}
\end{center}
\end{table}

Algorithm~\ref{qfgeo_algo} shows \QfGeo's handling of a flow that arrives at node $v$. In Lines 1 and 2, it first predicts the ellipse factor $\ell$ through the function $L$ that calculates $\LCap$ from Eq.~\ref{eq_ellipse_facotr_capacity} at either the source node or the relay hits the maximum retransmission times.  It then forwards packet $p$ within the ellipse with factor $\ell$. The actions for forwarding packet $p$ at $v$ are specified in Algorithm~\ref{qfgeo_forward}. Relevant notations are listed in Table~\ref{notationQFGeo}.

The first line in Algorithm~\ref{qfgeo_forward} filters the candidate forwarders from node $v$'s neighbors, which are located within the bounded region of $p$. Note that since \QfGeo\ applies a depth-first search, these candidates should be unvisited and excluded from the nodes that are backtracking to $v$.  In addition to satisfying the capacity requirement, the forwarding prioritizes the candidates with sufficient residual capacity $\Theta(u)$ for  flow $p.\!f\!l$. Lines 2 to 10 specify the $\epsilon$-greedy approach, which forwards packets when the set of candidates is not empty. Packet $p$ is sent immediately to the last forwarder of flow $p.\!f\!l$ with a probability of $1-\epsilon$. Otherwise, $v$ forwards $p$ to the node closest to $p$'s destination. Lines 11 to 16 handle the case where there is no qualified candidate filtered in Line 1; node $v$ still tries to find a neighbor with the best routing progress to $p.dst$. Finally, in Lines 17 to 19, if all of $v$'s neighbors within the forwarding region are already visited, $v$ must backtrack packet $p$ to its parent $u$. In that case, $u$ marks $v$ in $BL(p, u)$ so that $v$ will not be the prioritized forwarder in $p$'s associated flow.

\begin{algorithm}[t]
\SetAlgoLined
\If{$v = p.src \vee RETX(p, v) > RETX_{max}$}{
        $p.\ell := L(p.src, p.dst, \rho')$\;
}
FORWARD($p$, $v$, $\ell$)\;
\caption{\QfGeo($p ,v$)}
\label{qfgeo_algo}
\end{algorithm}

\begin{algorithm}[t]
\SetAlgoLined
$S := \{u |  u \! \in \! N\!E(p, v, \ell) \wedge u \! \not\in \! V(p) \wedge u \! \not\in \! BL(p, v) \wedge \Theta(u) \! \geq \! C(p.\!f\!l) \}$\;

\uIf{$S \neq \{\}$}{
    \eIf{$rnd < 1-\epsilon \wedge q(p.\!f\!l, v) \in S$}{
        forward $p$ to $u:=q(p.\!f\!l, v)$\;
    }{
        forward $p$ to $u:= \operatorname*{argmin}_{x\in S} Dist(x, p.dst)$\;
        $q(p.\!f\!l, v) := u$\;
    }
    $Parent(p, u) := v$\;
    $V(p) := V(p) \cup \{v\}$\;
}
\uElseIf{$S = \{\} \wedge \{u | u\in N\!E(p, v, \ell) \wedge u\not\in V(p) \} \neq \{\} $}{
    $S := \{u | u\in N\!E(p, v, \ell) \wedge j\not\in V(p)\}$\;
    forward $p$ to $u:= \operatorname*{argmin}_{x\in S} Dist(x, p.dst)$\;
    $q(p.\!f\!l, v) := u$\;
    $Parent(p, u) := v$\;
    $V(p) := V(p) \cup \{v\}$\;
}
\Else{
    backtrack to $u := Parent(p,v)$\;
    $BL(p, u) := BL(p, u) \cup {v}$\;
}
\caption{FORWARD($p, v, \ell$)}
\label{qfgeo_forward}
\end{algorithm}

\section{Evaluation Results}
\subsection{Configuration Space of Network Emulation}
Our validation of \QfGeo\ uses the ns3 simulator, which we extended in several ways: to support jammer nodes and to emulate our real-code implementation of the protocol.\footnote{We chose to develop a real-code implementation to facilitate translation-to-practice efforts in the incorporation of \QfGeo\ in recent SDR platforms that have been adopted by the DoD. The extension to support emulation was based on an integration of ns-3 with the Direct Code Execution framework \cite{NS3DCE}.}. The validation compares the performance of \QfGeo\ with that of greedy forwarding (GF) and maximum capacity routing (MCR). GF is the geographic routing that greedily forwards packets to the nodes closest to the destination. MCR is a reactive protocol that collects regional network states within an elliptic region predicted by Eq.~\ref{eq_prediction} to find the maximum capacity route for the entire flow transmission. To also calibrate the impact of bounded forwarding, we simulate two versions of \QfGeo: one using a bounded region (\QfGeo) and the other without a bounded region (\QfGeo-A). All routing protocols are tested with point-to-point flows and the MAC communication over CSMA.

In the networks we simulate, each node is equipped with two radios that share a fixed capacity of 8 Mbps. Each radio operates with one dedicated channel of 1 Mbps for the control plane and two channels of 3.5 Mbps for data plane communication; when serving more than two flows, each radio uses round-robin scheduling for the flows. We apply a time-slotted MAC communication with a 3 ms slot for the data plane. Every slot is long enough to transmit a 1000-byte data packet and receive an acknowledgment between a pair of nodes within some transmission radius.

With respect to interference, each radio is configured with the same transmission and interference radius of 1km ($\Delta=1$). And with respect to simulation scenarios with external interference, a single adversarial jammer with a +3 dB above the nominal transmission power and a 100\% effective jamming range of 1.32km is placed in the center of the network layout. Outside of this radius, jamming contributes to node-local RF noise conditions and could still cause blocking interference depending on the SNR margin of the potential victim link, just as any other simulated RF emitter would. The jammer continually generates jamming signals over one channel to reduce the network data capacity by 3.5Mbps (i.e., from 7Mbps to 3.5Mbps) within its interference range. The validation does so both for environments with and without external interference.

The configuration space of experiments we conducted is four-dimensional, namely: (i) network size, (ii) network density, (iii) number of concurrent flows, and (iv) mobility. With respect to these four dimensions, we simulated networks with nodes distributed over a rectangular region using a uniform random distribution placement model.  For (i), we let the network size range over 27, 64, 125, and 216; the results presented here emphasize the cases of 64 nodes and 125 nodes. For (ii), we let network density $\rho$ range over $\sqrt{2}$, 2, 3, 4, and 5. For (iii), we let the number of concurrently arriving flows range over 1, 3, 7, and 10. While we considered different flow sizes as well, for reasons of space, the results presented here are for the case of 3 Mbit per flow. And for (iv), we let the network be static or to be mobile; again, while we considered multiple speeds, the mobility result presented here is for the case of 10 m/s. 

Each of our experiment configurations consists of four trials of 30 seconds. Each trial starts with no assumption of each node's neighborhood or network state. To obtain the states of the neighborhood (e.g., coordinates and regional capacity $\Theta$), each node periodically broadcasts a Hello message over the control channel every 200 ms. All flows concurrently arrive after 5 seconds from the start of each trial. We note that each flow generates one packet at each slot. Once a flow arrives, \QfGeo\ and GF immediately start forwarding packets at the corresponding source nodes. In MCR, the arrival of flows at the source triggers route exploration to retrieve the state of the subgraph within a bounded ellipse, and then the source calculates the maximum capacity path for routing. At the end of each trial, the destination of a flow updates the received bits, flow time, and end-to-end latency for calculating the performance metrics.

In terms of the performance metrics, we compare the goodput, packet reception ratio, and average latency of different protocols. For the network configurations with concurrent flow transmission, we calculate an average result taken over the different numbers of flows with packets received at their destination. The goodput is defined as $\frac{total\ received\ bits}{average\ f\!low\ time}$, where the flow time of a single flow is the interval between the first sent packet at the source and the last received packet at the destination during the simulation time. Note that a high goodput can be resulted not only from a flow with reliable path(s) that transmit all packets but also from a flow with path(s) that deliver only a few packets in a short time. We thus propose the metric of goodput efficiency, which takes into account both the goodput and the reception ratio to evaluate the communication efficiency and reliability of a routing protocol in diverse scenarios. Goodput efficiency is defined as follows:
\begin{equation}
\begin{array}{l}
goodput\ e\!f\!\!f\!iciency = \frac{goodput \times {reception\ ratio}}{bandwidth \times \#f\!lows} \in [0, 1] ,
\label{eq_goodput_efficiency}
\end{array}
\end{equation}
where bandwidth represents the bit rate over the entire radio spectrum. Perfect goodput efficiency is achievable in flows with a one-hop path that is free from interference and collision.

\subsection{Performance in Static and Mobile Networks}
\begin{figure}[!t]
\minipage{0.236\textwidth}
  \subfloat[Goodput Efficiency, Size 125]{\label{fig_ge_static_unjamming_size125}{\includegraphics[width=\linewidth]{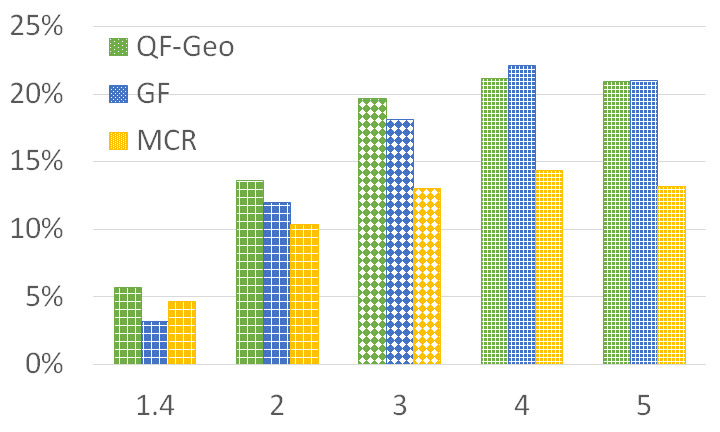}}}
\endminipage\hfill
\minipage{0.236\textwidth}%
  \subfloat[Goodput Efficiency, Size 64]{\label{fig_ge_static_unjamming_size64}{\includegraphics[width=\linewidth]{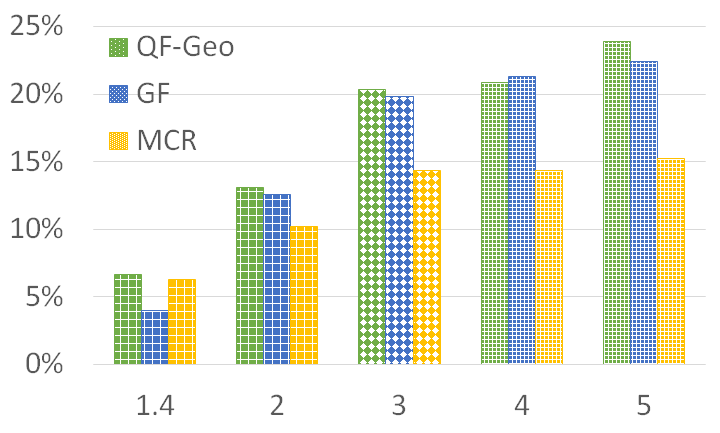}}}
\endminipage\hfill
\\
\minipage{0.236\textwidth}
  \subfloat[Packet Reception Ratio, Size 125]{\label{fig_rr_static_unjamming_size125}{\includegraphics[width=\linewidth]{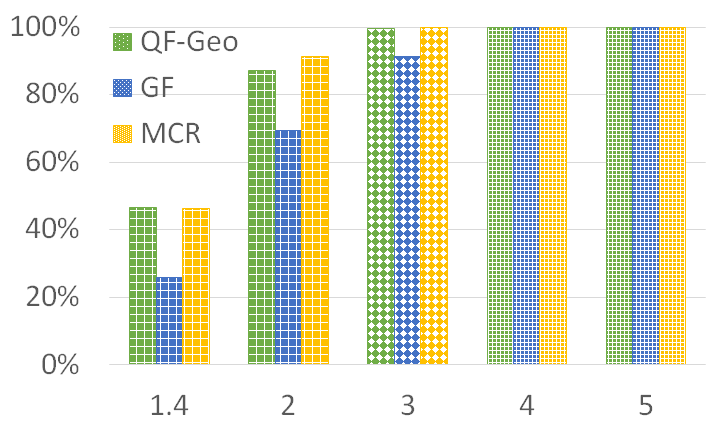}}}
\endminipage\hfill
\minipage{0.236\textwidth}%
  \subfloat[Packet Reception Ratio, Size 64]{\label{fig_rr_static_unjamming_size64}{\includegraphics[width=\linewidth]{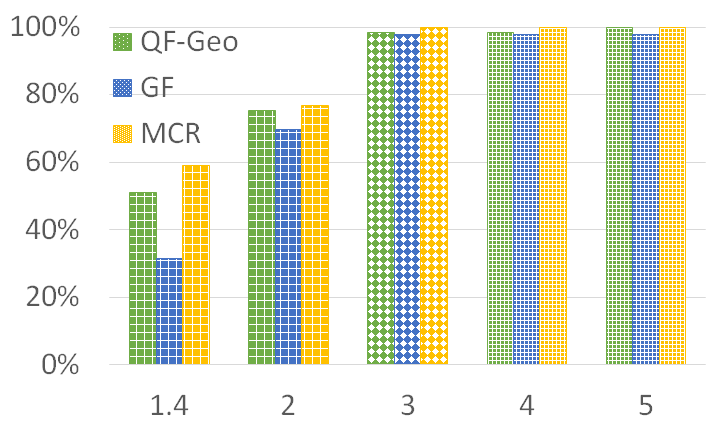}}}
\endminipage\hfill
\\
\minipage{0.236\textwidth}
  \subfloat[Latency (sec), Size 125]{\label{fig_l_static_unjamming_size125}{\includegraphics[width=\linewidth]{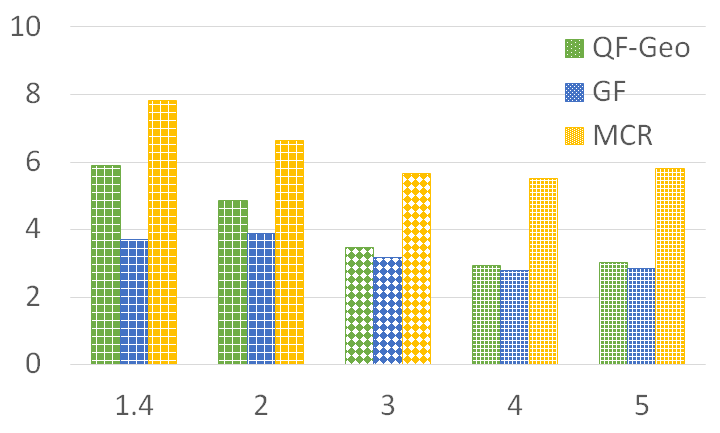}}}
\endminipage\hfill
\minipage{0.236\textwidth}%
  \subfloat[Latency (sec), Size 64]{\label{fig_l_static_unjamming_size64}{\includegraphics[width=\linewidth]{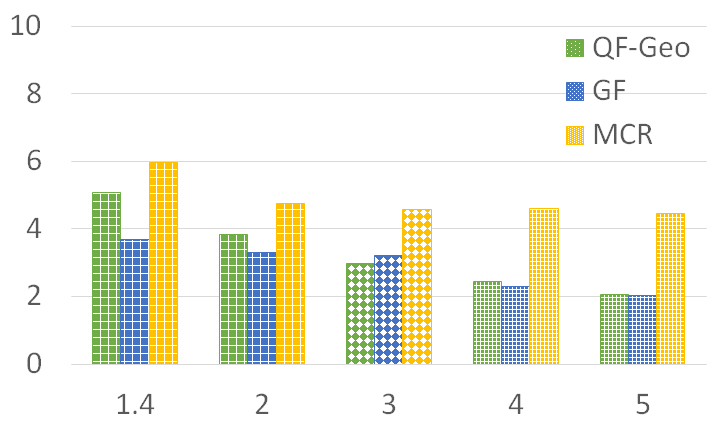}}}
\endminipage\hfill
\caption{Density versus different performance metrics for \QfGeo\ versus GF and MCR at networks with 125 and 64 static nodes. We use different fill patterns of columns to represent the three types of connected networks: large grids denote foam-like networks ($\rho \! = \! 1.4, 2$), diamond grids denote connected networks between foam-like and fog-like ones ($\rho \! = \! 3$), and small grids denote fog-like networks ($\rho \! = \! 4, 5$).  Each point-to-point flow is 3Mbit. Goodput efficiency and packet reception ratio are typically improved by \QfGeo; its latency is competitive, especially if its improved reception ratio in the presence of interference is accounted for.}
\label{fig_static_unjamming}
\end{figure}

\begin{figure}[!t]
\minipage{0.236\textwidth}
  \subfloat[Goodput Efficiency, Size 125]{\label{fig_ge_mobile_unjamming_size125}{\includegraphics[width=\linewidth]{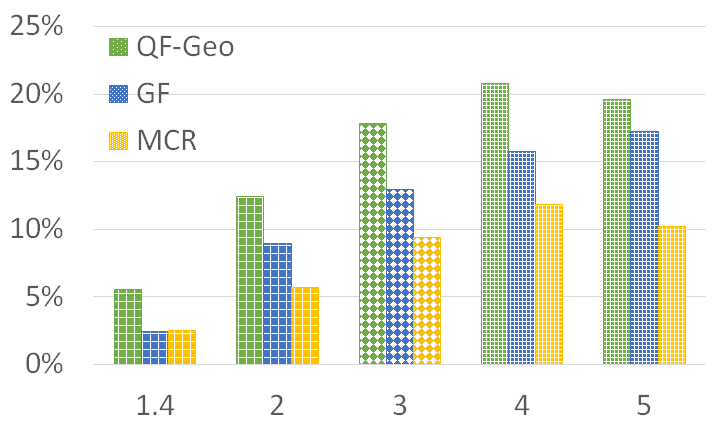}}}
\endminipage\hfill
\minipage{0.236\textwidth}%
  \subfloat[Goodput Efficiency, Size 64]{\label{fig_ge_mobile_unjamming_size64}{\includegraphics[width=\linewidth]{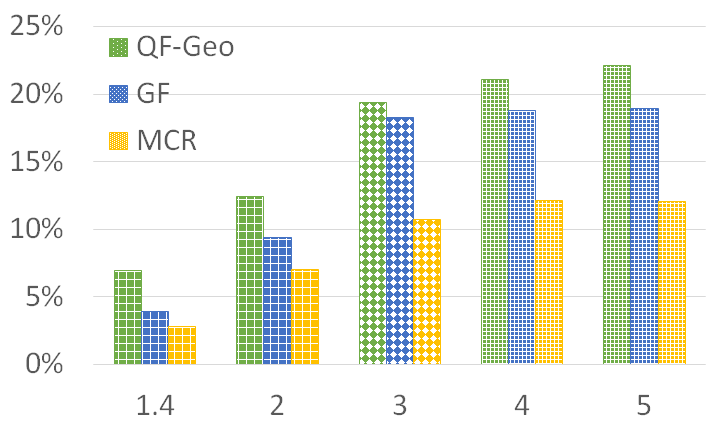}}}
\endminipage\hfill
\\
\minipage{0.236\textwidth}
  \subfloat[Packet Reception Ratio, Size 125]{\label{fig_rr_mobile_unjamming_size125}{\includegraphics[width=\linewidth]{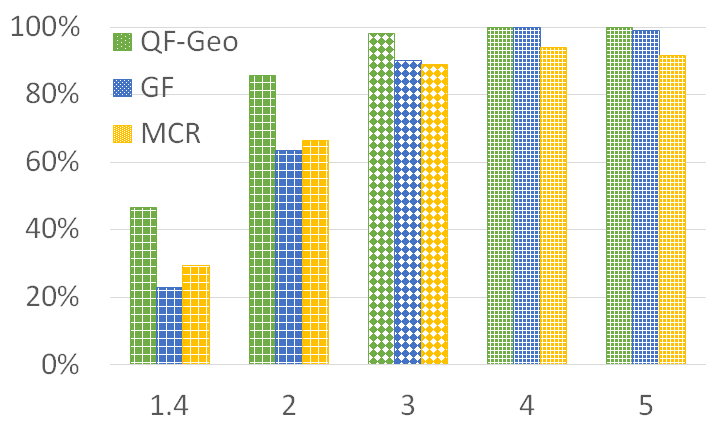}}}
\endminipage\hfill
\minipage{0.236\textwidth}%
  \subfloat[Packet Reception Ratio, Size 64]{\label{fig_rr_mobile_unjamming_size64}{\includegraphics[width=\linewidth]{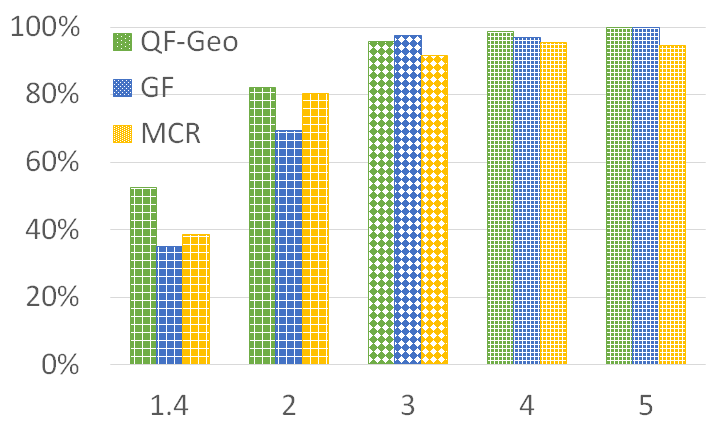}}}
\endminipage\hfill
\\
\minipage{0.236\textwidth}
  \subfloat[Latency (sec), Size 125]{\label{fig_l_mobile_unjamming_size125}{\includegraphics[width=\linewidth]{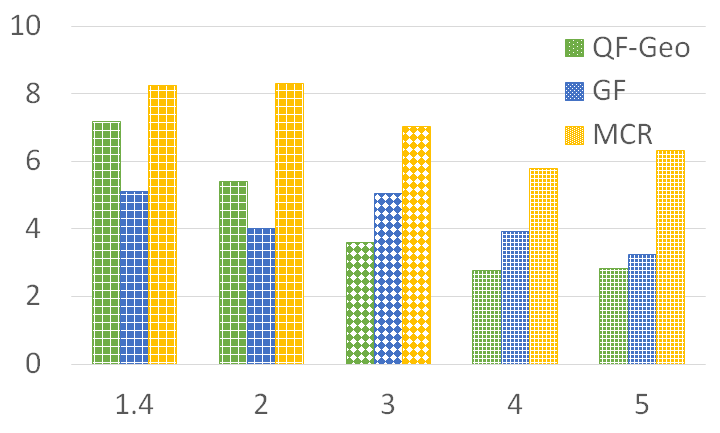}}}
\endminipage\hfill
\minipage{0.236\textwidth}%
  \subfloat[Latency (sec), Size 64]{\label{fig_l_mobile_unjamming_size64}{\includegraphics[width=\linewidth]{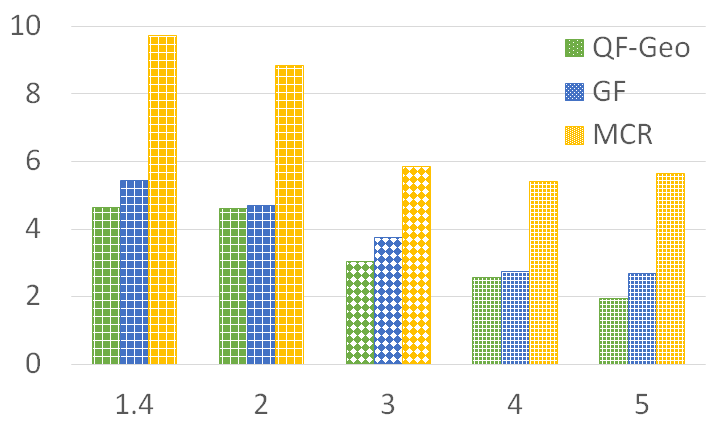}}}
\endminipage\hfill
\caption{Density versus different performance metrics for \QfGeo\ versus GF and MCR at networks with 125 and 64 mobile nodes. Goodput efficiency, reliability, and latency are typically improved by \QfGeo.}
\label{fig_mobile_unjamming}
\end{figure}

Figs.~\ref{fig_static_unjamming} and~\ref{fig_mobile_unjamming} show that \QfGeo\ works well over all configurations of density, size, and mobility. Note that the average source-destination distance reduces as the network size decreases from 125 to 64. This change reduces the average length of routing paths and may then yield lower latency.

In static network scenarios (Fig.~\ref{fig_static_unjamming}), although GF high goodput efficiency in fog-like networks ($\rho \! = \! 4, 5$) since its direct forwarding enjoys a high reception ratio and low latency from a dense node distribution. Yet, it underperforms in goodput efficiency and reception ratios in foam-like  (i.e., minimally connected such as with $\rho \! = \! \sqrt{2}$) networks; this low performance is due to failures in its handling of holes. It has good latencies overall, given its bias towards direct paths, but this comes at the expense of reliability. In contrast, MCR receives high goodput efficiency at density $\sqrt{2}$ but low goodput efficiency at densities of 2, 3, 4, and 5, because its average flow time includes a non-negligible portion of exploration time that significantly reduces the goodput efficiency.  Its reliability tends to be high because it searches for the best paths in its region of exploration.
The results of MCR imply that a snapshot-based protocol always incurs overhead for network state collection, even if that is within a bounded region. The resulting exploration time is insufficiently amortized over the data transmission time, especially in flows with a small amount of data, and the latencies are consistently high as well. Compared with both GF and MCR, \QfGeo\ achieves significantly better goodput efficiency and reliability in foam-like networks; and even in fog-like networks, it does consistently better than MCR and is competitive with GF. 
Overall, the results show that \QfGeo\ effectively generalizes to the scenarios over various densities in the static case.

More noticeably, in mobile network scenarios (Fig.~\ref{fig_mobile_unjamming}), across all configurations, \QfGeo\ substantially outperforms GF and MCR in terms of goodput efficiency and packet reception ratio and at least significantly outperform MCR in terms of latency. Importantly, there is no obvious gap in the performance between the static and mobile networks in \QfGeo. 
This is because its forwarder prioritization and memory of the last forwarder sustain the reliability of routing paths and the communication efficiency over moving nodes. Since a packet can be rapidly forwarded to the destination along the pre-explored route, \QfGeo\ efficiently reduces the chance of backtracking packets by avoiding frequent exploration of new routes. On the other hand, the path reliability in GF and MCR is undermined as nodes become mobile. Unlike \QfGeo's filtering out of underperforming relays, GF is prone to forward packets over long links, which are likely broken. Furthermore, MCR may suffer from the staleness of the  network state, causing it to choose unreliable routing paths.

\subsection{Performance with respect to Jamming}
\begin{figure}[!t]
\minipage{0.236\textwidth}
  \subfloat[Goodput Efficiency, Size 125]{\label{fig_ge_static_jamming_size125}{\includegraphics[width=\linewidth]{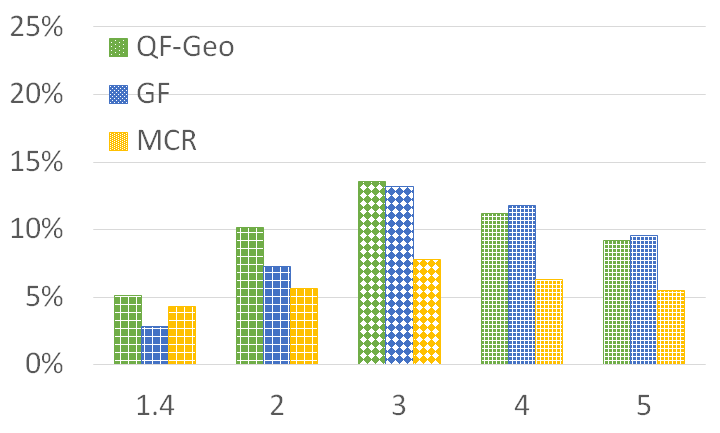}}}
\endminipage\hfill
\minipage{0.236\textwidth}%
  \subfloat[Goodput Efficiency, Size 64]{\label{fig_ge_static_jamming_size64}{\includegraphics[width=\linewidth]{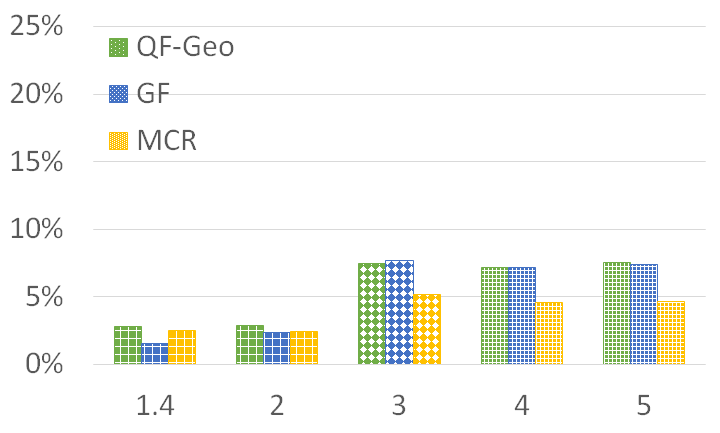}}}
\endminipage\hfill
\\
\minipage{0.236\textwidth}
  \subfloat[Packet Reception Ratio, Size 125]{\label{fig_rr_static_jamming_size125}{\includegraphics[width=\linewidth]{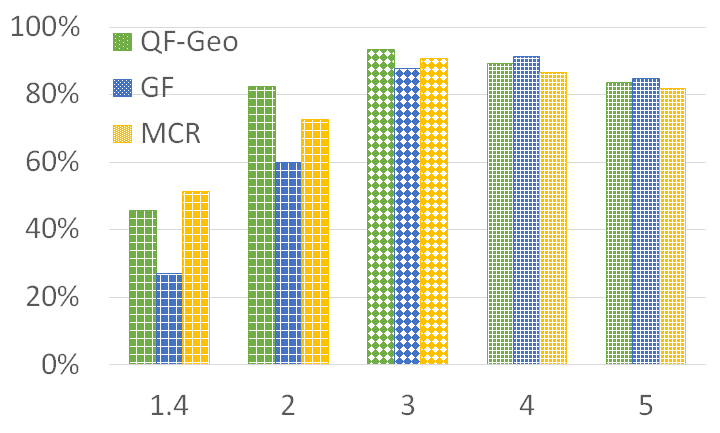}}}
\endminipage\hfill
\minipage{0.236\textwidth}%
  \subfloat[Packet Reception Ratio, Size 64]{\label{fig_rr_static_jamming_size64}{\includegraphics[width=\linewidth]{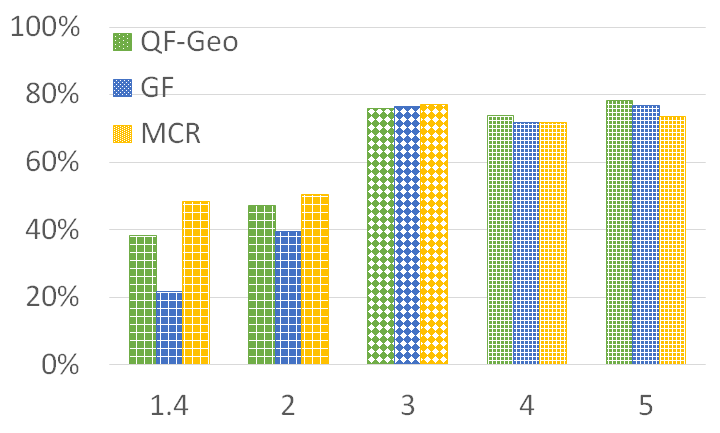}}}
\endminipage\hfill
\\
\minipage{0.236\textwidth}
  \subfloat[Latency (sec), Size 125]{\label{fig_l_static_jamming_size125}{\includegraphics[width=\linewidth]{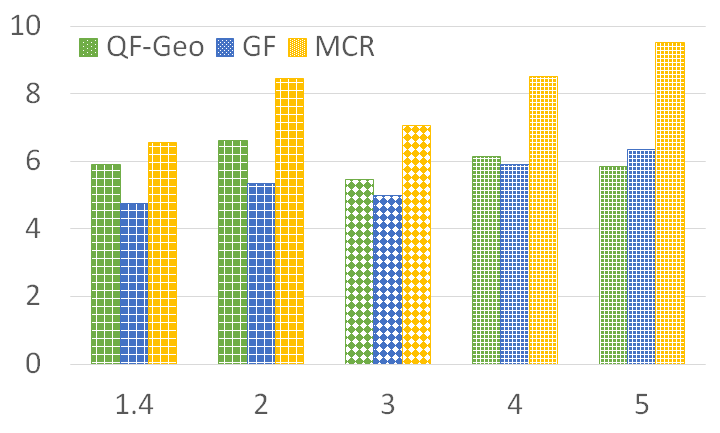}}}
\endminipage\hfill
\minipage{0.236\textwidth}%
  \subfloat[Latency (sec), Size 64]{\label{fig_l_static_jamming_size64}{\includegraphics[width=\linewidth]{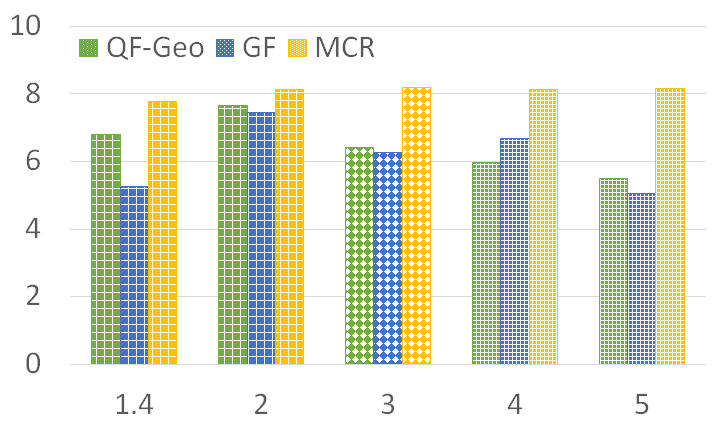}}}
\endminipage\hfill
\caption{Density versus different performance metrics for \QfGeo\ versus GF and MCR at networks with 125 and 64 static nodes and an additional jammer located at the center of network. Goodput efficiency and packet reception ratio are typically improved by \QfGeo; its latency is competitive, especially if its reception ratio in the presence of interference is accounted for.}
\label{fig_static_jamming}
\end{figure}

\begin{figure}[!t]
\minipage{0.236\textwidth}
  \subfloat[Goodput Efficiency, Size 125]{\label{fig_ge_mobile_jamming_size125}{\includegraphics[width=\linewidth]{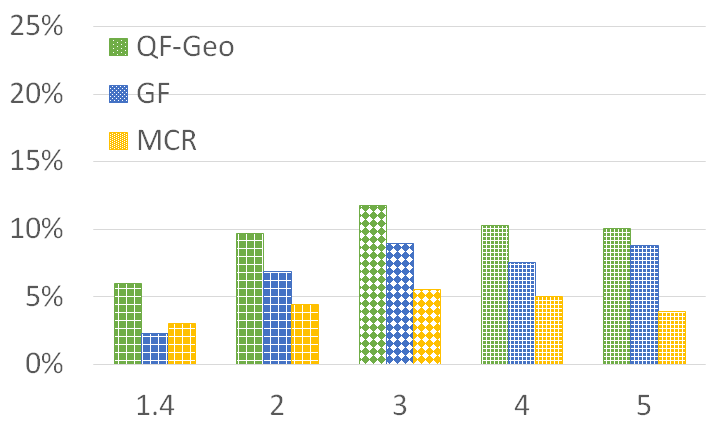}}}
\endminipage\hfill
\minipage{0.236\textwidth}%
  \subfloat[Goodput Efficiency, Size 64]{\label{fig_ge_mobile_jamming_size64}{\includegraphics[width=\linewidth]{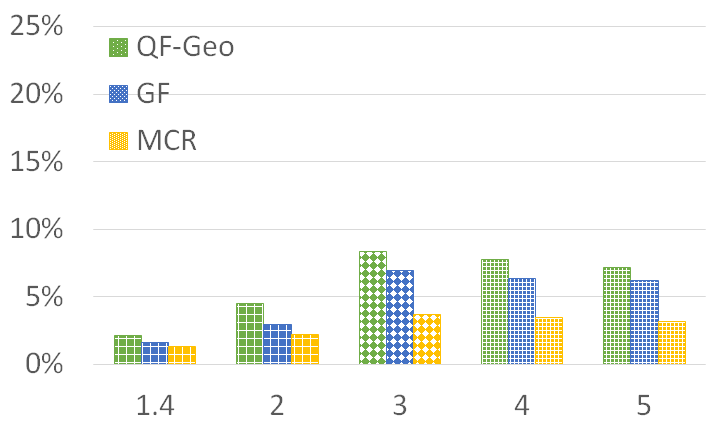}}}
\endminipage\hfill
\\
\minipage{0.236\textwidth}
  \subfloat[Packet Reception Ratio, Size 125]{\label{fig_rr_mobile_jamming_size125}{\includegraphics[width=\linewidth]{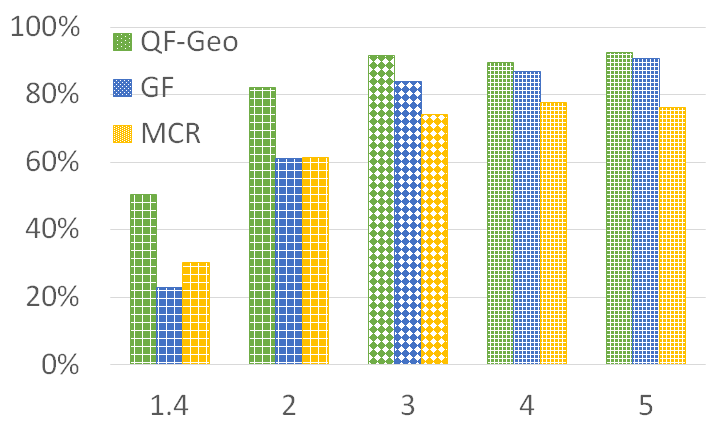}}}
\endminipage\hfill
\minipage{0.236\textwidth}%
  \subfloat[Packet Reception Ratio, Size 64]{\label{fig_rr_mobile_jamming_size64}{\includegraphics[width=\linewidth]{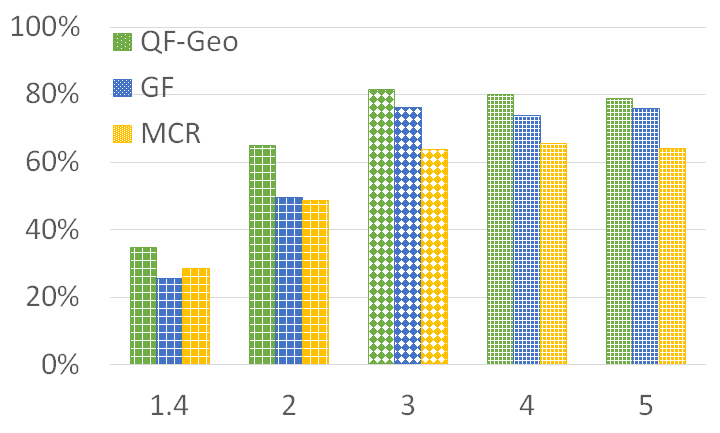}}}
\endminipage\hfill
\\
\minipage{0.236\textwidth}
  \subfloat[Latency (sec), Size 125]{\label{fig_l_mobile_jamming_size125}{\includegraphics[width=\linewidth]{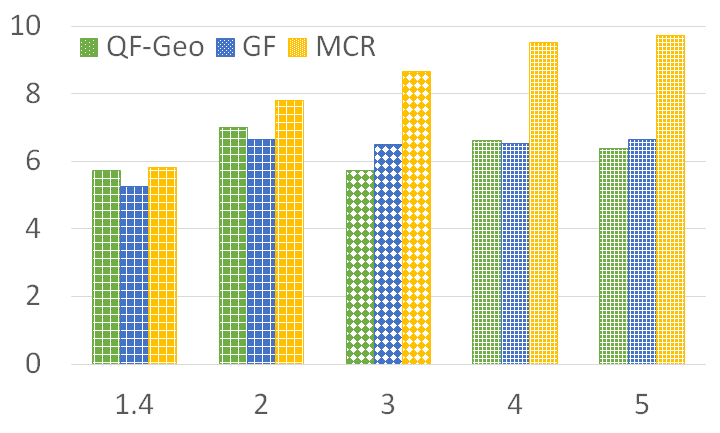}}}
\endminipage\hfill
\minipage{0.236\textwidth}%
  \subfloat[Latency (sec), Size 64]{\label{fig_l_mobile_jamming_size64}{\includegraphics[width=\linewidth]{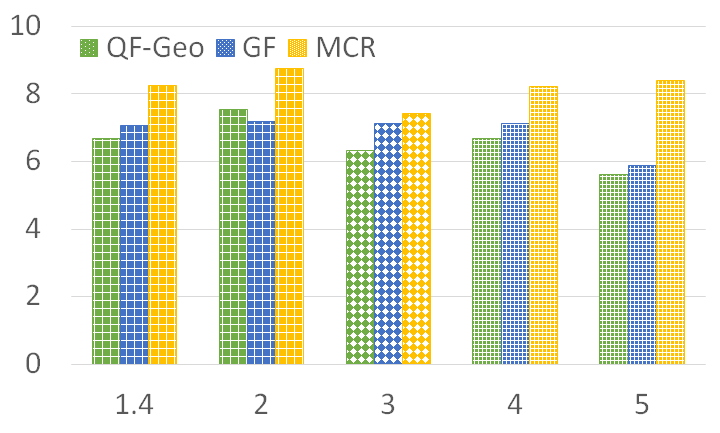}}}
\endminipage\hfill
\caption{Density versus different performance metrics for \QfGeo\ versus GF and MCR at networks with 125 and 64 mobile nodes and an additional jammer located at the center. Goodput efficiency, reliability, and latency are typically improved by \QfGeo.}
\label{fig_mobile_jamming}
\end{figure}

Figs.~\ref{fig_static_jamming} and~\ref{fig_mobile_jamming} show that \QfGeo\ effectively eschews the jammed region to achieve high goodput efficiency and packet reliability across varying densities, sizes, and mobility speeds. We note that as the network size decreases from 125 to 64, the number of flows interfered with by the jamming region increases, and then the performance is likely to degrade. 

In static network scenarios, \QfGeo\ consistently outperforms in terms of goodput efficiency and packet reliability in foam-like networks and remains competitive in fog-like networks. Compared to \QfGeo\ and GF, MCR does not achieve the highest reception ratio in high-density networks because its calculation of bounded region is not aware of the current jamming traffic. A failure to extend the ellipse area (and thus increase the space diversity) likely yields more interference as the degree of concurrent communication and the external jamming traffic increase. Therefore, \QfGeo\ achieves comparably good performance metrics in all density configurations due to the adaptation of ellipse size to both density and network capacity.

In mobile network scenarios, \QfGeo\ clearly outperforms GF and MCR in terms of goodput efficiency and reception ratio across all configurations; its latency is competitive with GF and better than that of MCR. \QfGeo's performance in mobile cases is sometimes even higher than that of \QfGeo\ in static cases; this is because mobility may offer additional chances to route packets away from jammed regions. 

\subsection{Impact of Bounded Forwarding on Improved Performance}

\begin{table*}[!htbp]
\centering
\caption{The improvement of performance metrics in \QfGeo\ compared to \QfGeo-A.}
\begin{tabular}{*8c}
\toprule
\multicolumn{2}{c}{Network Scenario} &  \multicolumn{2}{c}{Goodput Efficiency} & \multicolumn{2}{c}{Reception Ratio} & \multicolumn{2}{c}{Latency}\\
\midrule
{} & {} & w/o jammer & w/ jammer  & w/o jammer & w/ jammer & w/o jammer   & w/ jammer\\

\multirow{2}{*}{Static} & foam-like & \textcolor{TBcolor}{+0.93\%} & \textcolor{TBcolor}{+4.40\%} & \textcolor{TBcolor}{+0.78\%} & \textcolor{TBcolor}{+1.89\%} & \textcolor{TBcolor}{-0.04\%} & \textcolor{TBcolor}{-1.92\%}\\\cline{2-8}
 & fog-like & 0\% & \textcolor{TBcolor}{+0.14\%} & 0\% & \textcolor{TBcolor1}{-0.05\%} & 0\% & \textcolor{TBcolor}{-0.68\%}\\\hline
\multirow{2}{*}{Mobile} & foam-like & \textcolor{TBcolor1}{-1.12\%} & \textcolor{TBcolor}{+0.49\%} & \textcolor{TBcolor1}{-0.40\%} & \textcolor{TBcolor1}{-1.09\%} & \textcolor{TBcolor}{-1.50\%} & \textcolor{TBcolor}{-6.96\%}\\\cline{2-8}
 & fog-like & 0\% & \textcolor{TBcolor}{+0.05\%} & 0\% & \textcolor{TBcolor1}{-0.01\%} & 0\% & \textcolor{TBcolor}{-0.13\%}\\
\bottomrule
\end{tabular}
\label{tb_gfgeo_comparison}
\end{table*}

In this section, we compare two versions of \QfGeo\ , \QfGeo\ (with a bounded region) and \QfGeo-A (without a bounded region), to show the effect of forwarding packets within an ellipse whp including the shortest path. The results show on average bounded forwarding achieves comparable performance in terms of goodput efficiency and reception ratio but can further reduce the end-to-end latency. 

Table~\ref{tb_gfgeo_comparison} compares the difference in performance metrics across configurations of mobility and density between \QfGeo\ and \QfGeo-A. The green and red fonts respectively denote positive and negative percentage differences in the metrics between the two versions. On average, the difference in goodput efficiency and reception ratio is very slight, indicating that bounding is sufficient for capturing the efficiency. We note that in the scenarios of fog-like networks without jamming traffic, bounded forwarding yields an identical performance to the unbounded case because the routing paths with high capacity are geometrically direct and can be greedily found by both \QfGeo\ and \QfGeo-A. Also, the cases with increased reception ratio in \QfGeo\ imply that bounded forwarding may achieve more packet delivery ratio in a limited experiment period. However, it may also reduce the reception ratio in low-density networks since, with low probability, a routing path does not exist in a bounded region for some flows.

The results also show that on average and typically, \QfGeo\ has an improved latency across all configurations. In foam-like networks, \QfGeo\ shows a reduction of latency up to 1.50\% in cases without jamming traffic and 6.96\% in cases with jamming traffic by using a bounded forwarding scheme. We note that the cases with the most reduction of latency happen in the configurations of mobility and foam-like networks. This implies that using bounded forwarding can effectively reduce the length of routing paths in sparse networks, and the gain becomes more significant while frequently dealing with broken links in mobile networks.

\vspace*{-2mm}
\section{Conclusions and Future Work}


A probabilistic framework allows for a tighter bound on the region of exploration than a deterministic framework while only rarely failing to find a shortest path.

We analyzed the geometric region of forwarding consistent with high goodput efficiency and reliability and found that across the full range of densities, this region is well approximated by an ellipse parameterized by the source-destination distance and the density.  For foam-like scenarios, the region of sufficient exploration is a ``fat'' ellipse; and for fog-like networks, the region is a ``skiny'' elongated ellipse. This approach contrasts with most existing routing algorithms, which make implicit assumptions that limit their effectiveness to either fog-like or foam-like network densities. 

We presented a geographic routing protocol based on bounded exploration over these elliptical regions that are robust to network dynamics with densities, including mobility, external interference, and changes in traffic patterns.  Via network emulation, we experimentally validated that substantial improvements in goodput efficiency, reliability, and end-to-end latency are achieved at scales of up to several hundreds of nodes.


In the near term, our focus will be on refinements of the coordination with other network layers to provide distributed multi-channel control that achieves higher capacity-delay performance, using reinforcement learning techniques across layers.  We are also interested in learning-based approximations of \QfGeo\ and its refinements.  Moreover, our work has uncovered opportunities for further enhancements of SDNs and 5G networks with a hybrid wired-wireless architecture.

\bibliographystyle{IEEEtran}
\bibliography{myref}
    
\end{document}